\documentclass[aps,prb,amsmath,amsfonts,floatfix,superscriptaddress,color,twocolumn]{revtex4-1}

\usepackage{graphicx}
\usepackage{amsmath,amssymb,amsfonts}
\usepackage{gensymb}
\usepackage{bm}
\usepackage{upgreek}
\usepackage{wasysym}
\usepackage{bibentry}
\usepackage[T1]{fontenc}
\usepackage{textcomp}
\usepackage[scaled=0.92]{helvet}
\usepackage{courier}
\usepackage{color}
\usepackage{leftidx}   

\usepackage{bbold} 
\usepackage{relsize}

\usepackage[english]{babel}
\usepackage[latin1]{inputenc}

\usepackage{ifthen}
\usepackage{footnote}
\usepackage[normal]{subfigure}

\newcommand{\be}{\begin{equation}}
\newcommand{\ee}{\end{equation}}
\newcommand{\ba}{\begin{align}}
\newcommand{\ea}{\end{align}}
\newcommand{\sysb}{\left\{\begin{array}}
\newcommand{\syse}{\end{array}\right.}
\newcommand{\baa}{\begin{array}}
\newcommand{\eaa}{\end{array}}
\newcommand{\bs}{\begin{split}}
\newcommand{\es}{\end{split}}

\newcommand{\matb}{\left(\begin{array}}
\newcommand{\mate}{\end{array}\right)}

\newcommand{\mal}{\mathcal}
\newcommand{\rmd}{{\rm{d}}}

\newcommand{\rme}[1]{{\rm{e}}^{#1}}
\newcommand{\mand}{\quad\text{ and }\quad}

\newcommand{\wh}{\widehat}

\newcommand{\ob}{\mal{O}}
\newcommand{\id}{\mathbb{1}}
\newcommand{\trace}[1]{{\rm tr}\left\{ #1 \right\}}

\newcommand{\gh}{\phantom}

\newcommand{\ha}{\frac{1}{2}}
\newcommand{\eq}{\, = \,}
\newcommand{\set}[1]{\left\{  #1 \right\}  }

\newcommand{\pop}[1]{\hat{n}_{#1}}

\newcommand{\hsi}{\wh{\sigma}}
\newcommand{\whN}{\wh{N}}
\newcommand{\quadblock}[4]{\matb{c|c} #1 & #2 \\ \hline #3 & #4 \mate}
\newcommand{\bE}{{\bf E}}

\newcommand{\lt}{\left(}
\newcommand{\rt}{\right)}
\newcommand{\lqq}{\left[}
\newcommand{\rqq}{\right]}
\newcommand{\lan}{\left\langle}
\newcommand{\ran}{\right\rangle}
\newcommand{\abs}[1]{\left| #1 \right|}

\newcommand{\av}[1]{\lan #1 \ran}

\newcommand{\sz}{{\sigma^z}}
\newcommand{\sxM}{\matb{cc} 0 & 1 \\ 1 & 0   \mate}
\newcommand{\syM}{\matb{cc} 0 & -i \\ i & 0   \mate}
\newcommand{\szM}{\matb{cc} 1 & 0 \\ 0 & -1   \mate}

\newcommand{\R}{\mathbb{R}}
\newcommand{\N}{\mathbb{N}}
\newcommand{\Z}{\mathbb{Z}}

\newcommand{\ket}[1]{\left| #1 \ran}
\newcommand{\bra}[1]{\lan #1 \right|}

\newcommand{\comm}[2]{\left[ #1, #2 \right]}
\newcommand{\acomm}[2]{\left\{ #1, #2 \right\}}

\newcommand{\sina}[1]{\sin \left(  #1 \right)}
\newcommand{\tana}[1]{\tan \left(  #1 \right)}

\newcommand{\sinna}[2]{\sin^{#1} \left(  #2 \right)}

\newcommand{\reff}[1]{(\ref{#1})}

\newcommand{\prodl}[2]{\prod\limits_{#1}^{#2}}
\newcommand{\suml}[2]{\sum\limits_{#1}^{#2}}

\newcommand{\liml}[1]{\lim\limits_{#1}}

\newcommand{\twopartsdef}[4]
{
	\left\{
		\begin{array}{ll}
			#1 & \mbox{if } #2 \\
			#3 & \mbox{if } #4
		\end{array}
	\right.
}

\newcommand{\figwid}{0.45}

\begin{document}
\title{Prethermalization from a low-density Holstein-Primakoff expansion}

\author{M. Marcuzzi}
\affiliation{School of Physics and Astronomy, University of Nottingham, Nottingham, NG7 2RD, UK}
\author{J. Marino}
\affiliation{Institut f\"ur Theoretische Physik, Universit\"at zu K\"oln, D-50937 Cologne, Germany}
\author{A. Gambassi}
\affiliation{SISSA --- International School for Advanced Studies, via Bonomea 265, 34136 Trieste, Italy}
\affiliation{INFN --- Istituto Nazionale di Fisica Nucleare, sezione di Trieste}%
\author{A. Silva}
\affiliation{SISSA --- International School for Advanced Studies, via Bonomea 265, 34136 Trieste, Italy}

\date{\today}

\begin{abstract}
We consider the non-equilibrium dynamics arising after a  quench of the transverse magnetic field of a quantum Ising chain, together with the sudden switch-on of a long-range interaction term. The dynamics  after the quantum  quench is mapped onto a fully-connected model of hard-core bosons,
after a suitable combination of a Holstein-Primakoff transformation and of a low-density expansion in the quasi-particles injected by the quench. This mapping holds for a broad class of initial states and for quenches which do not cross the   critical point of the transverse field Ising model.  We then study the algebraic relaxation in time of a number of observables towards a metastable, pre-thermal state,  which becomes the asymptotic steady state in the thermodynamic limit.

\end{abstract}


\maketitle

\begin{section}{Introduction}

The constant progress in manipulating cold atomic gases has provided  insight into the non-equilibrium dynamics of isolated, interacting quantum many-body systems. The experimental observation of the absence of relaxation in an (almost) integrable one-dimensional Bose gas\cite{Kinoshita} and the appearance of an intermediate, metastable regime in the dynamics of a non-integrable system on time-scales much shorter than those required for its equilibration \cite{Preth-exp1,Preth-exp2,Preth-exp3,Preth-exp4, Jorg15, Langen2016} call for a better understanding of the mechanisms underlying quantum relaxation.

The inherent unitarity of the evolution makes the emergence of relaxation, thermalization and, more generally, decoherence a subtle issue, since they cannot really occur in isolated systems\cite{NonTh}: in fact, a unitary dynamics cannot turn an initial pure state into a thermal distribution, which is a proper statistical mixture. Consequently, one may look for signs of thermalization in a suitable set of local observables \cite{FagottiReduced}, whose expectations, in the long-time limit, are equivalently captured by the thermal distribution -- its temperature being determined by the average energy of the initial state \cite{VonNeumann,Nonloc-th,Loc-nonloc}. 

In order to probe the relaxation of quantum many-body systems, the protocol known as \emph{quantum quench} has provided a convenient conceptual framework, widely employed in recent investigations \cite{Silva, Eisert}. It consists in suddenly changing a parameter of the many-body Hamiltonian (magnetic field, interaction strength, or coupling among spins, etc) on time scales short enough to leave the system initially frozen in the ground state of the pre-quench Hamiltonian. This state will then evolve according to the post-quench Hamiltonian: since in general it does no longer constitute an energy eigenstate, in the case of a global quench it will have a finite energy density above its ground state, and hence a finite density of quasi-particles, whenever they are defined. 

While one can think of the thermal ensemble as the natural candidate for the stationary state of an isolated system \cite{Q-GGE1,Kollath,ETH3,Santos,Biroli,Th-Mitra1,TavoraMitra,Rigol2,Lux}, there are instances in which the presence of conservation laws increases the amount of information preserved during the course of the evolution. In particular, \emph{integrable} systems are characterized by an extensive amount of independent integrals of motion, which are at the root of their exact solvability. Accordingly, one can construct the corresponding maximal entropy state\cite{GGE-Jaynes}, which is usually referred as the \emph{generalised Gibbs ensemble }(GGE), and which is effectively approached in the long-time limit \cite {Cazalilla, RigolGGE, Cramer, Barthel, Fioretto, Caux, DynaEssler, ColluraSot, QAction, Brockmann, Pozsgay, CollCal, DeNardis, SotCal, Panfil, Prosen, Essler2016, Calabrese2016, Caux2016}. 

A richer scenario is expected to emerge when the quench involves two parameters with competing effects, i.e., one preserving and  one spoiling the integrable nature of the quantum system. For instance, if a weak integrability-breaking interaction is switched on, while simultaneously quenching a control parameter which in the absence of interactions would preserve  the integrability of the model,  the intermediate time evolution is expected to be still approximately determined by the integrable (non-interacting) part of the Hamiltonian \cite{Kollar}. As a consequence, a two-stage relaxation emerges: First, the system approaches the GGE corresponding to the integrable part of its Hamiltonian, such that this GGE approximately describes a metastable state at intermediate times after the quench. Afterwards, on time scales which depend on the strength of the interaction, the effect of quasi-particle scattering becomes relevant and drives the system towards an eventual, \emph{bona fide} thermalization. This mechanism is usually referred to as \emph{prethermalization} (or as \emph{prerelaxation} when, instead, the perturbation breaks non-abelian integrability into integrability \cite{Fagotti14}), and, although this notion was firstly employed in the context of high-energy quantum field theories \cite{Preth1}, during the last few years it has been extended to the domain of condensed matter physics. Signatures of prethermalization have been reported in long-range interacting quantum simulators \cite{Worm}, in systems of spinless fermions with tunable interactions \cite{Robinson, Nessi, Bertini15}, in two-dimensional spinless Fermi gases \cite{Iucci}, in metastable superfluids \cite{Rosch},  in the three-dimensional isotropic Heisenberg model \cite{Demler}, for interaction quenches in the fermionic Hubbard model \cite{Werner, Stark, Kehrein, Kehrein2}, during aging in interacting $\phi^4$-theories (either isolated or in contact with a bath)\cite{Gagel2014, Alessio, Maraga2015, AlessioLong}, in noisy Ising models \cite{Jamir1, MarinoLong}, and in interacting Luttinger liquids \cite{Buchhold, MitraPreth, Ueda}.

In this work, we investigate in more detail  the prethermalization dynamics of the long-range model we originally introduced in Ref.~\onlinecite{M-preth}.  
Starting from the quench of the quantum Ising chain, we introduce a long-range spin-spin interaction which breaks many, but not all, of the original conservation laws, as we detail in Sec.~\ref{sec:model}. We first show in Sec.~\ref{ssec:mapping} that an exact mapping exists to a model of hard-core bosons on a fully-connected lattice. As long as the quasi-particle density generated by the quantum quench remains sufficiently small, one can think of the hard-core constraint as being substantially ineffective, and thus treat the bosons as ordinary ones. Formally, this corresponds to a lowest-order truncation of the Holstein-Primakoff transformation \cite{HP}. This approximation -- which holds for small quenches up to very large times (cfr with Fig.~\ref{fig:compI}), renders the theory effectively non-interacting and allows us to map the non-equilibrium dynamics of the original  model onto the relaxation of an integrable one.  
We should notice  that a similar low-density expansion in the quasi-particles produced by the quench has already been employed for studying the relaxation dynamics of isolated interacting quantum systems in different contexts, and including both integrable \cite{Nonloc-th,Loc-nonloc} and non-integrable systems \cite{Lux, Calabrese16}.

We then proceed to solve numerically the dynamics of the approximately-equivalent bosonic model, highlighting the presence of plateaus in the relaxation of some physically relevant observables, which are typically approached algebraically in time; the main numerical results are reported in Sec.~\ref{sec:num}. In Sec.~\ref{sec:validity} (which extends the results of \cite{M-preth}) we discuss in more detail the range of applicability of our approach and offer some numerical evidence that it provides reasonably accurate results even for quenches at the critical point or across it, in spite of the fact that they generate highly-populated modes and hence violate the low-density condition. 

Contrary to several previous studies on prethermalization plateaux\cite{Robinson, Nessi, Iucci, Kehrein, Kehrein2, Stark, MitraPreth}, our analysis does not rely on a straightforward perturbative expansion in the parameter controlling the interaction among quasi-particles of the pre-quench integrable model and offers a new perspective into the study of non-equilibrium dynamics of interacting systems, which is in general a formidable analytic challenge.

The model presented here is exactly solvable in the thermodynamic limit, since the long-range nature of the interaction suppresses fluctuations and makes mean-field theory exact (see Section \ref{sec:model}). This fact has been employed in Ref.~\onlinecite{fagottibertini} in order to solve, \emph{inter alia}, the time evolution of this model through a mean-field mapping.%
%

Despite the diversity of approaches developed in order to tackle the quench dynamics of this system, we think it is worth  reporting here in full detail our approach, since it suggests that in certain cases pre-thermalization in interacting systems can be understood as the dynamics of an approximate integrable model, which emerges at intermediate time scales and which is not  perturbatively connected to the pre-quench model. 


	
\end{section}	
	
	

	\begin{section}{The model and the quench protocol}
	\label{sec:model}
	
The interacting spin model under study, with Hamiltonian $H = H_0 + V$, is a quantum Ising chain of $N$ sites with periodic boundary conditions, where
\be
	H_0(g)=- \frac{J}{2} \suml{i=1}{N} \left( \hsi_i^x \, \hsi_{i+1}^x + g\hsi_i^z \right),
\ee
is the standard short-range Ising model, with $J$ the nearest-neighbor coupling and $Jg$ the transverse magnetic field, while $\hsi_i^{x, y, z}$ denote the standard spin-$\tfrac{1}{2}$ operators acting on site $i$, which can be represented as Pauli matrices (see Eq.~\eqref{eq:Pauli}). This $H_0$ is perturbed by a long-range interaction of the form
\be \label{interaction}
		V= \frac{ \lambda}{N}  \lt \frac{ M^z - \overline{M^z} }{2} \rt^2,
\ee
where $M^z\equiv \sum_i \hsi_i^z$ is the global transverse magnetization and $\overline{M^z}$ its time average calculated according to
\be
	\overline{M^z} = \liml{T\to \infty} \frac{1}{T} \int_0^T \rmd t \,\, \rme{i H_0 t} M^z \rme{-i H_0 t}.
	\label{eq:Mz_bar}
\ee
Note that, throughout the paper, we set $\hbar = 1$ and measure energies and inverse times in units of $J$, thereby fixing $J=1$. 
    

%

In order to drive this system out of equilibrium, we consider here a composite quench protocol: We ``prepare'' the system in the ground state $\ket{\psi_0 (g_0)}$ of the pre-quench Hamiltonian $H_0 (g_0)$. At $t=0$, we vary the magnetic field ($g_0 \to g$) on a time scale so short that the state is effectively left unaffected, simultaneously switching on the interaction term $V$. In other words,
\be
	g(t) = \twopartsdef{g_0}{t<0}{g}{t \geq 0}  \mand   \lambda(t) = \twopartsdef{0}{t<0}{\lambda}{t\geq 0.}
\ee	
Accordingly, the dynamics for $t \geq 0$ takes place under the action of the post-quench Hamiltonian, i.e.,
\be
	\ket{\psi(t)} = \rme{-i(H_0 (g)+V) t} \ket{\psi_0 (g_0)}.
\ee
The Hamiltonian $H_0(g)$ is known to be integrable \cite{QPT, QQIsing1, QQIsing2, CalabrFagott} and can be mapped via a sequence of Jordan-Wigner\cite{JW}, Fourier and Bogoliubov transformation to a free model of fermionic quasi-particles $\gamma_k$, $\gamma_k^\dag$, $k$ being the quasi-momentum or reciprocal wave vector (for a brief summary of this approach we refer to Appendix \ref{ssec:QIsing}). The integrability  of $H_0$ is however spoiled in the presence of the many-body interaction $V$  which generate scattering among the quasi-particles and allows a redistribution of momentum and energy among the different mode populations $\pop{k} = \gamma_k^\dag \gamma_k$.


Note that switching on the interaction $V$ at time $t=0$ without the quench  $g_0\to g$, would be sufficient to make the ground state $\ket{\psi_0 (g_0)}$ of the pre-quench Hamiltonian a non-equilibrium one for $H$; however, the quench $g_0\to g$ in the magnetic field has the effect of providing a non-vanishing quasi-particle density from the very beginning of the evolution; in particular, introducing the shorthand $\ket{0} = \ket{\psi_0(g_0)}$, one obtains
	\be
		{ \bra{0} \pop{k}  \ket{0}} \equiv  \bra{0} \gamma_k^\dag  \gamma_k  \ket{0} = \sinna{2}{\theta_k(g) - \theta_k(g_0)},
	\label{eq:nk0}
	\ee
where $\theta_k (g)$ and $\theta_k(g_0)$ are the Bogoliubov angles corresponding to $H_0(g)$ and $H_0(g_0)$, respectively, and can be determined from Eq.~\eqref{eq:bangle0}, while $\pop{k} = \pop{k} (g)$ is the aforementioned mode population of the post-quench Ising Hamiltonian $H_0(g)$. Hereafter, all quantities are calculated in terms of the post-quench fermionic basis, unless otherwise stated (e.g., $\theta_k$ is shorthand for $\theta_k(g)$). 
Some examples of post-quench populations $\pop{k} (t \to 0^+)$ are displayed in Fig.~\ref{fig:pop} for various values of the initial and final transverse fields $g_0$ and $g$. 
		\begin{figure*}[!ht]
		\centering
		\includegraphics[width = \textwidth, trim = 5mm 0 45mm 0]{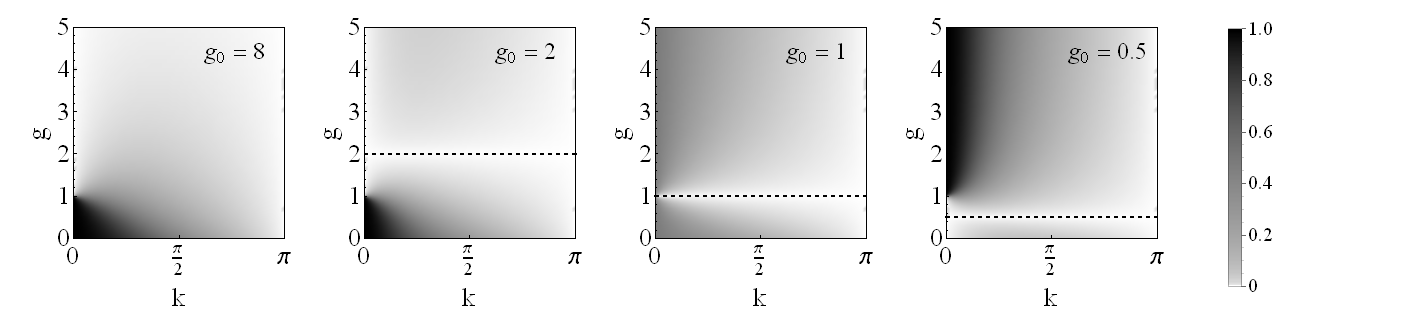}  
		\caption{ (\emph{colour online}) Post-quench population of the Ising quasi-particle modes $\pop{k}$ (cfr. Eq.~\eqref{eq:pop1}) at time $t = 0^+$ as a function of the final transverse field $g$ and the momentum $k$ for four different choices of the pre-quench field, from left to right $g_0 = 8$, $2$, $1$ and $0.5$. The dashed lines correspond to $g = g_0$, i.e., the no-quench case, in which the initial fermionic populations vanish. Ising modes become significantly populated only for quenches which cross from the ferromagnetic  ($g<1$) to the paramagnetic ($g>1$)  phase or viceversa, or start or end at the critical point $g=1$. Note that these plots are invariant under the exchange $g_0 \leftrightarrow g$, since $\pop{k}$ has the same property (see Eq.~\eqref{eq:nk0}). 
	}
		\label{fig:pop}
		\end{figure*}

Before proceeding, we comment on the subtraction of $\overline{M^z}$ in Eq.~\eqref{interaction}, which is the long-time average value of the global transverse magnetization in a quenched non-interacting Ising model ($\lambda=0$). The reason for this choice (further discussed in Appendix \ref{subtrcapp}) is grounded in the non-equilibrium dynamics of the connected correlation function $\av{M^z (t_1) M^z (t_2)}_c = \av{M^z (t_1) M^z (t_2)} - \av{M^z (t_1)}\av{M^z (t_2)}$ of the global transverse magnetization $M^z$
for which (see also Refs. \onlinecite{Foini, Foini2})
	\be
		\liml{\tau \to \infty} \,\liml{t \to \infty} \,\lan M^z(t+\tau) \, M^z (t) \ran_c > 0.
	\label{eq:clusterp}
	\ee
This fact indicates the presence of information which is never really lost, as measurements of this observable separated by an arbitrary time $\tau$ are still correlated. The ``cluster property'' can be restored also for this non-local observable with the subtraction of $\overline{M}^z$, which practically corresponds to subtract \emph{ab initio} from the interaction term $V$ a term which is extensive in the system size ($\sim O(N)$) and quadratic in the Ising integrals of motion $n_k$. 

\begin{subsection}{Remaining conserved quantities}

The integrability of the Ising model is typically expressed in terms of the conservation of the quasi-particle densities $\pop{k}$. This breaks down with the introduction of the interaction $V$. However, one can still identify an extensive number of conserved quantities. Their presence is more easily highlighted in the fermionic representation of the model (see Appendix \ref{ssec:QIsing}), in which $H_0$ reads
	\be
		H_0 (g) =  \suml{k > 0}{} \epsilon_k \psi_k^\dag \sigma^z \psi_k\,,
	\label{eq:IsingH1}
	\ee
where 
	\be
		\psi_k = \matb{l} \gamma_k\\ \gamma^\dag_{-k} \mate \quad \text{ and } \quad \psi^\dag_k = \matb{l} \gamma^\dag_k\\ \gamma_{-k} \mate^\intercal ,
		\label{eq:spinor}
	\ee
are Nambu spinors constructed  with the fermionic annihilation (creation) operators $\gamma_k$ ($\gamma_k^\dag$), while 
	\be
		\epsilon_k\equiv\sqrt{1+g^2  -  2 g\cos k }
	\label{eq:disprel0}
	\ee
is the dispersion relation of the quasi-particles. The sum runs over the discrete values $k = 2\pi (n-1/2) / N$ with $n = 1 \ldots N/2$, while $\sigma^{x, y, z}$ denote the usual Pauli matrices reported for convenience in Eq. \eqref{eq:Pauli}.
In the same basis, the interaction term $V$ is  in the form
%
%
%
	\begin{equation}
		\label{eq:hambog}
		V= \frac{\lambda}{N}\Big[\suml{k>0}{} \sin (2\theta_k) \; \psi_k^\dag\sigma^y\psi_k\Big]^2  .
	\end{equation}
In the notation of Eq.~\eqref{eq:hambog} the long-range nature of $V$ becomes apparent, as the terms $\psi^\dag_k \psi_k \psi^\dag_q \psi_q$ clearly connect every possible pair of momenta $k , q$. The total Hamiltonian $H $ is thus
	\be
		H = \suml{k>0 }{} \epsilon_k \, \psi_k^\dag \sz \psi_k + \frac{\lambda}{N} \lqq \suml{k>0}{} \sin (2\theta_k) \, \psi_k^\dag \sigma^y \psi_k \rqq^2.
	\label{eq:FH}
	\ee

 	By exploiting the identity $\comm{A^2}{B} = A \comm{A}{B} + \comm{A}{B}A$ and the fact that $\comm{H_0(g)}{\pop{k}} = 0$, one can show that
 	\begin{widetext}
	\be
	\begin{split}
		\comm{H}{\hat{n}_k} = \comm{V}{\hat{n}_k} = i \frac{\lambda}{2N}   \acomm{M^z - \overline{M}^z \,}{ \,   \sina{2\theta_k} \lt \gamma_k^\dag \gamma_{-k}^\dag + \gamma_{-k} \gamma_k \rt } ,
	\end{split}
	\label{eq:commpop}
	\ee
	\end{widetext}
	which proves that, for any finite $N$, most populations are not conserved. The only exceptions are the two extremal cases $k = 0$ and  $k=\pi$, which however are present only in the thermodynamic limit $N \to \infty$ and can therefore be safely disregarded.	
We notice now that both the Bogoliubov angle $\theta_k$ and the two-particle operators $\gamma_k^\dag \gamma_{-k}^\dag$ and $\gamma_{-k} \gamma_k$ appearing on the r.h.s.~of Eq.~\eqref{eq:commpop} are odd under the inversion of the momentum $k \to -k$, i.e., $\gamma_k \gamma_{-k} = - \gamma_{-k}\gamma_{k}$. This implies $\comm{H}{\pop{k}} = \comm{H}{\pop{-k}}$, and therefore
	\be
		I_k = \pop{k} - \pop{-k}  \quad \quad (k > 0)
		\label{eq:charges}
	\ee
	commutes with the total post-quench Hamiltonian $H$ (similar operators depending on fermionic pairs at opposite momenta have been discussed also in  Ref.~\onlinecite{QQIsing-finite}).

\end{subsection}

\end{section}

\begin{section}{Mapping to hard-core bosons }
\label{ssec:mapping}
		Thanks to the set of  $N/2$ mutually commuting constants of motion $I_k$ (with $I_{-k} = -I_k$), the spin chain described by $H$ can be exactly mapped onto a quadratic (yet non-diagonal) Hamiltonian of hard-core bosons, as we now proceed to show, completing the brief discussion of Ref.~\onlinecite{M-preth}. 
		
We start by analysing the structure of the Hilbert space in the presence of these constraints: first of all, each $I_k$ involves a pair of modes with opposite momenta $k$ and $-k$. Consequently, it acts non-trivially only on the four-dimensional subspace spanned by the vectors $\ket{\emptyset_k}$, $\ket{k} = \gamma_k^\dag \ket{\emptyset_k}$, $\ket{-k} = \gamma_{-k}^\dag \ket{\emptyset_k}$ and $\ket{k,-k} = \gamma_k^\dag \gamma_{-k}^\dag \ket{\emptyset_k}$, where the vacuum $\ket{\emptyset_k}$ is annihilated by both $\gamma_k$ and $\gamma_{-k}$. From the definition \eqref{eq:charges} it is not too difficult to see that
\be
	I_k \ket{\pm k} = \pm \ket{\pm k} \mand I_k \ket{\emptyset_k} = I_k \ket{k, -k} = 0,
\ee
which identifies the possible eigenvalues $\left\{ -1, 0, 1 \right\}$ of each $I_k$. Note that $0$ is doubly degenerate, while $\pm 1$ have no degeneracy. Since the $I_k$'s commute, one can independently fix their eigenvalues for each $k > 0$, e.g.
\be
	\left\{ 1,0,0, \ldots  \right\} \equiv \left\{ I_{\pi/N} = 1,I_{3\pi/N}=0, I_{5\pi/N} = 0, \ldots  \right\}
	\label{eq:label}
\ee
indicates a list of possible eigenvalues of $I_k$, ordered with increasing allowed values of $k=(2n+1)\pi/N$ on the lattice, starting from $n=0$.
Each choice uniquely identifies a subspace (sector) of the Hilbert space which is orthogonal to the others and is not dynamically connected to them (i.e., the Hamiltonian $H$ acquires a block-diagonal structure). For instance, take a chain of $N=4$ spins; this has four quasi-momenta $\pm \pi/4$ and $\pm 3 \pi/4$ and, consequently, two conserved quantities $I_{\pi/4}$ and $I_{3\pi/4}$. The Hilbert space fragments into $9$ dynamically-disconnected sectors such as $\set{1,1}$ --- corresponding to the sole vector $\ket{\pi/4} \otimes \ket{3\pi/4}$ --- or $\set{0,-1}$ which is spanned, instead, by the two vectors  $\ket{\emptyset_{\pi/4}} \otimes \ket{-3\pi/4}$ and $\ket{\pi/4,-\pi/4} \otimes \ket{-3\pi/4}$.

%
%
%
%
%
%
%
%
%
%
%
%
%
%
%

Labels such as \eqref{eq:label} contain $N/2$ distinct eigenvalues, hence the total number of sectors is $3^{N/2}$. As mentioned above, the ``$0$''s are doubly-degenerate, implying that the actual dimension of a sector is $2^{N_0}$, with $N_0$ the total number of ``$0$''s appearing in the corresponding label. Considering still the  $N=4$ example reported above, the sector $\set{1,1}$ has no zeroes, so $N_0 = 0$ and is one-dimensional, while  $\set{0,-1}$  has instead $N_0 = 1$ and the sector is two-dimensional.  
Recalling that the Ising $\Z_2$ symmetry $\hsi^x \to - \hsi^x$ in the spin formulation corresponds to the preservation of the parity $(-1)^{\sum_k \pop{k}}$ of the fermion number, we also see that the number $ N/2 - N_0$ of $\pm 1$s in a label determines the $\Z_2$-parity of the sector, as it counts the number of unpaired quasi-particles present. For example, for $N = 8$, the sector $\left\{  0,0,1,-1\right\}$ has $N_0 = 2$ and is thus four-dimensional, while $N/2 - N_0 = 2$ is even, implying that all its vectors transform trivially under the $\Z_2$ transformation ($U_{\Z_2} \ket{v} = \ket{v}$, $\forall \ket{v} \in \left\{  0,0,1,-1\right\}$). Conversely, the sector $\left\{0,1,1,-1  \right\}$ is two-dimensional ($N_0 = 1$) and $N/2 - N_0 = 3$ is odd, which means that its vectors pick up a factor $-1$ instead ($U_{\Z_2} \ket{v} = - \ket{v}$, $\forall \ket{v} \in \left\{  0,1,1,-1\right\}$).
%
%
By construction, each sector also carries definite momenta $kI_k$ corresponding to each $(k,-k)$ subspace. 
		
We shall now look for operators which leave all sectors invariant; by the properties stated above, those operators must preserve both parity (they can only change the number of fermions by an even amount) and momentum components (they must carry zero net momentum in each ($k,-k$) subspace).
%
%
For any fixed choice of $k$, it is sufficient to consider only combinations of the basic creation and annihilation operators $\gamma_{\pm k}$, as all the others do not act on the corresponding subspace. The only (non-trivial) choices satisfying all constraints are the quadratic operators
		\be
			\pop{\pm k} = \gamma_{\pm k}^\dag \gamma_{\pm k},  \quad\quad b_k^\dag = \gamma_{k}^\dag \gamma_{-k}^\dag,   \quad\quad   b_k = \gamma_{-k} \gamma_k\, ,
		\label{eq:opquad}
		\ee 
		which represent the populations and the creation and annihilation of pairs with zero net momentum, respectively, and the quartic one
		\be
			\pop{k}\pop{-k} = b_k^\dag b_k.
		\label{eq:opquart}
		\ee
		All other possible operators can be re-expressed in terms of these ones by making use of the canonical anticommutation relations  of the fermionic operators, see \reff{eq:canacomm}. Products of the operators above at equal or different momenta (e.g., $\pop{k} b^\dag_q b_p$) will leave all sectors invariant.

Every operator which commutes with all $I_k$'s can be expressed in terms of (products of) the ones in Eqs.~\eqref{eq:opquad} and \eqref{eq:opquart} and the Hamiltonian $H$ constitutes no exception. As a matter of fact, it can be written in terms of the sole ``pair'' operators as 
		\be
			\sysb{l}
				H =  \mathlarger{\suml{k>0}{}} \left[\epsilon_k - \displaystyle{\frac{\lambda}{N}} \sin^2(2\theta_k)\right] \left( I^2_k - 1 \right) + H'   , \\[7mm]
				H' = \mathlarger{\suml{k,q>0}{}} \left[2\beta_{kq} b_k^\dag b_q - \alpha_{kq}  (b_k^\dag b_q^\dag + b_k b_q)\right],
			\syse
		\label{eq:chernabog}
		\ee
		with
		\begin{subequations}
		\begin{align}
			\alpha_{kq} &= \frac{\lambda}{N} (1-\delta_{kq})\sina{2\theta_k} \sina{2\theta_q} , \label{seq:alpha}\\
			\beta_{kq} &= \epsilon_k \delta_{kq} + \alpha_{kq}. \label{seq:beta}
		\end{align}
			\label{eq:bmatrices}
		\end{subequations}
This implies that the relevant dynamics is described by the interaction of pairs of quasi-particles with zero net momentum, rather than of single fermionic modes, and that we can therefore reformulate the problem in terms of these new fundamental modes. In order to do so, we shall first investigate their nature. Since they obey
		\be
		\begin{split}
			\comm{b_k^{\dag}}{b_q^{\dag}} = \comm{b_k}{b_q} =  0 \ \ \ \  \mbox{with}\quad  \, k\neq q, \\  \acomm{b_k^{\dag}}{b_k^{\dag}} = \acomm{b_k^{\gh{\dag}}}{b_k^{\gh{\dag}}} =  0,  \quad \acomm{b_k}{b_k^\dag}  & = 1 - I_k^2,
		\label{eq:almostcomm}
		\end{split}
		\ee
		they behave almost, but not exactly, as hard-core bosons. In fact, this would require the last anticommutator above to be  $1$. On the other hand, by noticing that both $b_k$ and $b_k^\dag$ act as the null operator in a sector with $I_k = \pm 1$ (i.e., $b_k \ket{\pm k} = b_k^\dag \ket{\pm k} = 0$), we can effectively expunge them from $H'$. This operation leaves behind only those pair operators corresponding to momenta $q$ for which $I_q = 0$; within the corresponding eigensector they then satisfy the hard-core constraint.
		Thereby, in a sector characterized by having $N/2 - N_0$ unpaired quasi-particles, the projected Hamiltonian effectively describes a fully-connected model of hard-core bosons on a lattice with $N_0$ sites. The corresponding basis can be obtained by setting, for every involved $k$, the correspondence $\ket{\emptyset_k} \to \ket{{\bf 0}_k}$, $\ket{k,-k} \to \ket{{\bf 1}_k}$, where ${\bf 0}$ and ${\bf 1}$ stand for the boson being absent or present, respectively. 

		This reinterpretation clarifies  the effect  of introducing $V$ in the Hamiltonian of the Ising model: with $N/2$ conserved quantities still remaining, in fact, we cannot expect that the resulting model is completely non-integrable and, indeed, we identify sectors in which it is trivially solvable, which are the ones almost completely lacking pairs (i.e., those whose labels display just a few $0$s). For example, the $2^{N/2}$ totally-unpaired sectors collectively represent the zero-energy eigenspace of the Hamiltonian $H$ and coincide with the corresponding one of $H_0$; furthermore, each of the $(N/2) \times 2^{N/2-1}$ sectors having a single pair is two-dimensional and the corresponding reduced Hamiltonian is already cast in the diagonal form
		\be
			\matb{cc}
				-\epsilon_k + \mathlarger{\frac{\lambda}{N}} \sinna{2}{2\theta_k} & 0 \\
				0 & \epsilon_k + \mathlarger{\frac{\lambda}{N}} \sinna{2}{2\theta_k}
			\mate
		\ee 
		 in the basis $\left\{ \ket{{\bf 0}_k}, \ket{{\bf 1}_k}   \right\}$ introduced above. This is due to the presence of an additional symmetry, namely the conservation of the parity $\rme{i\pi \sum_k b_k^\dag b_k}$ of pairs, i.e.,
%
%
		\be
			\comm{H}{\rme{i\pi \sum_k b_k^\dag b_k}} = 0,
		\ee
		which further splits each sector in two halves of equal dimension. From a physical point of view, this is associated with the fact that $H'$ in Eq.~\reff{eq:chernabog} either does not affect their total number, $\sum_k b_k^\dag b_k$, or it simultaneously creates or destroys two pairs. 
		Although the structure of the states space becomes progressively more complicated as $N_0$ grows, it is clear that the model cannot really display non-integrable features as long as the dynamics remains confined in the small-$N_0$ sectors.

		The situation is reversed for $N_0 \approx N/2 \gg 1$;  although the corresponding eigensectors are exponentially smaller than the global Hilbert space (whose dimension is $2^N$), their dimensions are still exponentially large in the number of sites, as expected for a truly many-body problem. Note that, in spite of the fact that the Hamiltonian \reff{eq:chernabog} is quadratic in the pair operators, it does not define a free theory, due to the hard-core nature of the bosons; indeed, trying to diagonalise it by applying a generic Bogoliubov rotation	
		\be
			b_k = A_{kq} b_q' + B_{kq} b_q'^\dag 
		\ee
		intertwines the mixed commutation/anticommutation relations in Eq.~\reff{eq:almostcomm}, implying the impossibility of interpreting $b_k'$ and $b_k'^\dag$ as annihilation/creation operators of any kind of particles.
This relates to the fact that hard-core bosons are intrinsically interacting particles, as they can be thought as ordinary bosons subject to infinite ``on-site'' interparticle repulsion.

		Within our setting, the dynamics always starts from the totally-paired sector $N_0 = N/2$, independently of the values of the quench parameters $g_0$, $g$ and $\lambda$. The pre-quench state $\ket{0}$, in fact, can be represented as\cite{QQIsing1, Foini}
		\be
			\ket{0} = \ket{\psi_0 (g_0)} \propto \rme{i \sum_{k>0} t_k b^\dag_k}  \ket{\psi_0 (g)},
		\label{eq:istat}
		\ee
		where $t_k = \tana{\theta_k(g) - \theta_k(g_0)}$ and the effect of the operator on the r.h.s.~is to generate pairs on the vacuum of the post-quench Hamiltonian. Thereby, this whole class of initial states constitutes a suitable choice for highlighting the effects of $V$ on the dynamics, as it never involves the highly-unpaired sectors. 
		
		\subsection{Low-density approximation}
		
For $N \gg 1$, the interacting problem is hard to solve; in this case, instead of employing the usual perturbative expansion in the interaction strength, one can adopt a different approach, which makes use of the quadratic structure of the Hamiltonian. The point is that the hard-core constraint is expected to become effective only when the occupation of a given mode approaches $1$; as long as the quasi-particle densities remain much smaller than that, they behave approximately as standard bosons.
This condition must be first satisfied by the initial state. By inspecting Fig.~\ref{fig:pop} one sees that some modes get occupied to a relevant degree only in the case of quenches that connect different phases ($g > 1$ and $g_0 < 1$ or vice-versa) or that start ($g_0 \approx 1$) or end ($g \approx 1$) close to the critical point. Accordingly, we expect this approximation to be accurate for quenches within the same phase.

		 From a formal point of view, the pair operators $b_k$ and $b_k^\dag$ introduced in Eq. \eqref{eq:opquad} can be expressed in terms of standard bosonic ones $a_k$ and $a_k^\dag$ by means of a Holstein-Primakoff transformation\cite{HP}
		\be
			b_k = \sqrt{\id-a^\dag_k a_k} \, a_k \quad \text{ and } \quad b_k^\dag = a_k^\dag \sqrt{\id-a^\dag_k a_k};
		\ee
	for low densities,	one can then think of expanding the square roots as power series of their arguments, with
		\be
		\begin{split}
			b_k = \lt \id - \frac{1}{2} a^\dag_k a_k - \frac{1}{8}  (a^\dag_k a_k)^2 + \ldots \rt a_k , \\[2mm]
			b_k^\dag = a^\dag_k \lt \id - \frac{1}{2}a^\dag_k a_k - \frac{1}{8}  (a^\dag_k a_k)^2 + \ldots \rt,
		\label{eq:HPb}
		\end{split}
		\ee
		where $ a^\dag_k a_k$ is the bosonic number operator.
		Hence, as long as the average and the fluctuations of $a^\dag_k a_k$ remain small, one can conveniently truncate the expansions above. Further details are provided in App.~\ref{app:HP}.
What makes this approach particularly convenient is that, by expanding at the lowest order $b^\dag_k \approx a^\dag_k$ and $b_k \approx a_k$, one obtains in each sector a quadratic Hamiltonian which can now be diagonalised by a Bogoliubov rotation. Denoting by $K_S$ the set of paired momenta present in a given sector $S$ (see also note \footnote{For instance, if $S = \set{0,1}$ then $K_S = \set{\pi/4}$}), the expression of the reduced Hamiltonian acting on $S$ is
		\be
			\sysb{l}
				H_S = -\mathlarger{\suml{k \in K_S}{}} \left[\epsilon_k - \displaystyle{\frac{\lambda}{N}} \sin^2(2\theta_k)\right]  + H'_S ,  \\[4mm]
				H'_S = \mathlarger{\suml{k,q \in K_S}{}} \left[2\beta_{kq} a_k^\dag a_q - \alpha_{kq}  (a_k^\dag a_q^\dag + a_k a_q)\right].
			\syse
		\label{eq:chernabog1}
		\ee
		Note that the hard-core constraint $b_k^2 = b_k^{\dag 2} =0$ is reflected in the vanishing diagonal part of the matrix $\alpha_{kq}$ (see Eq.~\reff{seq:alpha}), which prevents terms such as $a_k^2 $ from appearing in $H_S$.

		As we show in the following, for most choices of the initial state --- as long as $g$ is not too close to the critical point $g_c = 1$ ---  the approximation above successfully describes  the early stages of the dynamics.
		At longer times, however, the quadratic approximation, encoded in the Hamiltonian \reff{eq:chernabog1}, is spoiled because higher-order terms in the expansion of the Holstein-Primakoff representation \reff{eq:HPb} become important, introducing interactions among the bosons. 

	\end{section}

\begin{section}{Pre-thermalized regime}
	\label{sec:num}
		
	With the notation of the previous Section, the ground state $|\psi_0(g_0)\rangle$ of $H_0(g_0)$ lies in the totally-paired sector of $H$, corresponding to the string of eigenvalues $\left\{ 000 \ldots 0 \right\}$. Thereby, the dynamics of any (combination) of the invariant quantities listed in Eqs.~\reff{eq:opquad} and \reff{eq:opquart} can be entirely determined within this subspace. Of course, working with a free bosonic Hamiltonian such as Eq. \reff{eq:chernabog1} comes with the substantial advantage of reducing the many-body problem to a one-body one; in other words, it is not necessary to diagonalise the full operator on the whole eigensector, but it is sufficient to determine the one-particle spectrum by applying an appropriate Bogoliubov transformation
		\be
		\begin{split}
			a_k =\sum_{q>0} \lqq  A_{k,q} \eta_q + B_{k,q} \eta_q^\dag   \rqq,  \\
			 a_k^\dag =  \sum_{q>0} \lqq A_{k,q}^\ast \eta_q^\dag + B_{k,q}^\ast \eta_q \rqq,
		\label{eq:Bogoliubov1}
		\end{split}		
		\ee
		which casts $H'$ in the diagonal form
		\be
			H' = \suml{q>0}{} E_q \eta^\dag_q \eta_q + \mathcal{C}, 
		\label{eq:freeH'}
		\ee
		where $\left\{ E_q \right\}_q$ are the energies of the bosonic modes (obtained through exact numerical diagonalization) and $\mathcal{C}$ is an incosenquential constant.	
		This problem amounts to the diagonalization of a $N \times N$ matrix, and is thus of polynomial complexity in $N$. As a consequence, one can perform a numerical analysis up to rather large system sizes. Details about the diagonalization procedure and the numerical computation of the relevant observables are provided in Appendix ~\ref{app:numdiag}.

		\begin{figure}[ht]
		\centering
		\includegraphics[width = 0.91 \columnwidth]{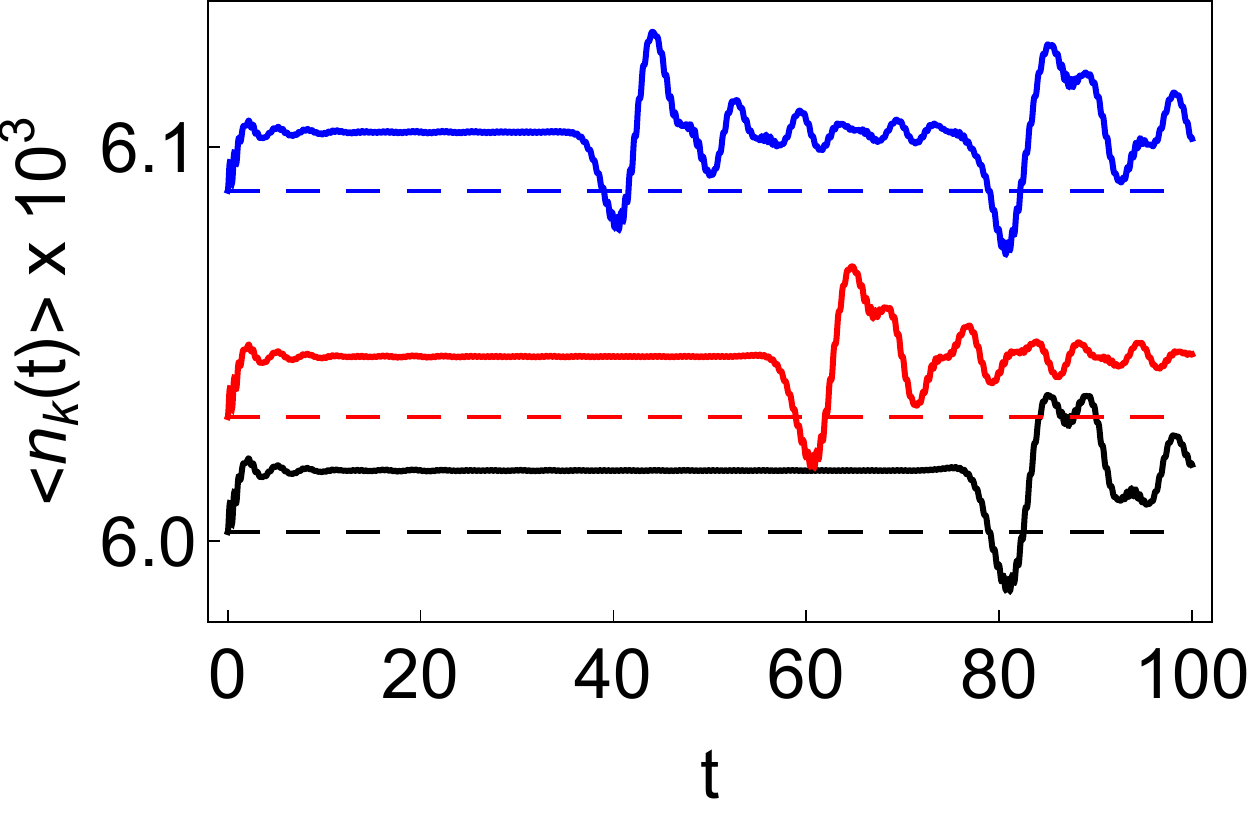}  
		\caption{(\emph{colour online}) Temporal evolution of $\av{\pop{k}(t)}$ for $k=\pi/2$, and a quench from $g_0 = 8$, to $g = 3$ and $\lambda = 1$, for various system sizes $N$: From top to bottom the solid lines correspond to $N=80, 120$ and $160$. The dashed lines of the respective colors indicate  the initial ($t=0$) value of each curve. These curves demonstrate that the population $\av{\pop{k}(t)}$ relaxes towards a prethermal value which is different from the initial one. The plateaus last until a recurrence time $t_R$, approximately equal to $N/2$. $t_R$ marks the reappearance of oscillations and it is due to the finite size of the chain.  
		}
		\label{fig:popdyn}
		\end{figure}
%

		By truncating the power series \reff{eq:HPb} derived from the Holstein-Primakoff representation to its leading terms, we are actually neglecting any form of interaction, as we end up with an Hamiltonian that is  non-interacting in each sector. On the other hand, taking into account next-to-leading orders clearly produces the appearance of higher-than-quadratic terms in Eq. \reff{eq:chernabog1}, which therefore give rise to quasi-particle interactions. In  light of this fact, we must conclude that there can be no sign of thermalization as long as the low-density approximation captures the physics of the system. In other words, if the system eventually leaves the prethermal regime highlighted here, it must do so on time scales longer than the regime of validity of the truncation which, as shown below, may encompass quite long time intervals. Prethermal features may only be observed within this regime, as it represents a stage at which  interactions among quasi-particles have not yet become dominant, leaving the system observables to relax on the quasi-stationary values associated to pre-thermalization. The typical prethermal behavior and time scales are illustrated in Fig.~\ref{fig:popdyn}, where representative plots are reported of the temporal evolution  of the central mode population $\av{\pop{\pi/2}}$ for various values of the system size $N$. 
		Quite clearly, after a short initial transient (of duration  $t\sim10$, independently of $N$), well-defined plateau values are attained in all cases, which define an intermediate, metastable stage of the evolution. As mentioned above, however, the dynamics of observables such as the occupations $\av{\pop{k}}$ can be decomposed in a finite collection of modes oscillating with some specific frequencies; accordingly, at finite size, the destructive interference which gives rise to the aforementioned plateaus cannot last indefinitely and, indeed, oscillations reappear at a recurrence time $t_R \simeq N/2$.

		\begin{figure*}[ht!]
		\centering

		\parbox{\figwid \textwidth}{
		\subfigure[]{\includegraphics[width = \figwid \textwidth]{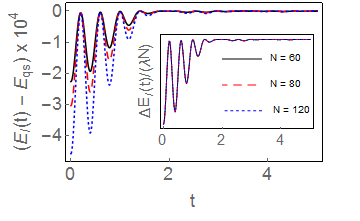} \label{sfig:EcollN} }}
			\hspace{1 mm}
		\parbox{\figwid \textwidth}{
		\subfigure[]{\includegraphics[width = \figwid \textwidth]{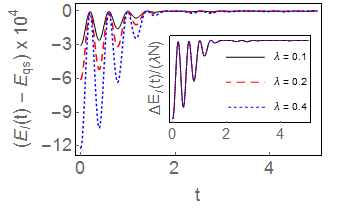} \label{sfig:EcollL} }} \\

		\parbox{\figwid \textwidth}{
		\subfigure[]{\includegraphics[width = \figwid \textwidth]{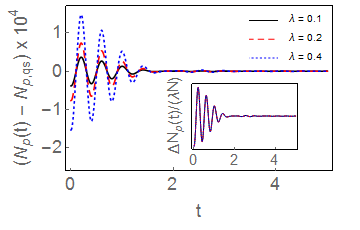} \label{sfig:NcollL}} }
			\hspace{1 mm}
		\parbox{\figwid \textwidth}{
		\subfigure[]{\includegraphics[width = \figwid \textwidth]{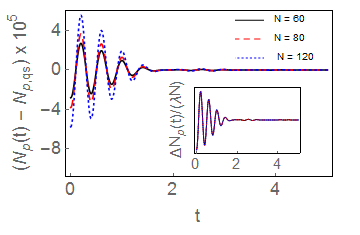} \label{sfig:NcollN}} } 
		\caption{ (\emph{colour online}) (a) Temporal evolution of the difference $\Delta E_I (t) = E_I(t) - E_{qs}$ between the Ising energy, $E_I$ (see Eq. \eqref{eq:E_0}), and its  stationary value, $E_{qs}$ after a quench from $g_0 = 9$ to $g=4$ with $\lambda = 0.1$, for three different system sizes $N=60$ (black), $80$ (red), and $120$ (blue). (b) Same quantities as in panel (a) at fixed $N = 80$ and for three different interaction strengths $\lambda = 0.1$ (black), $0.2$ (red), $0.4$ (blue). The inset of the various panels show the evolution of the same quantities as those displayed in the corresponding main plots, after a rescaling by a factor $1/(\lambda N)$, which makes the curves on each panel  collapse on a single master curve. (c) and (d): Evolution of the difference $\Delta N_p(t) = N_p(t) - N_{p,qs}$ (see Eq. \eqref{eq:forplot}) for exactly the same parameters as those used in (b) and (a), respectively. The insets are again rescaled by $1/(\lambda N)$ and  display collapse on a single master curve. 
}
		\label{fig:ENcoll}
		\end{figure*}


The pre-thermal behavior highlighted above for single populations is also reflected in the evolution of extensive observables such as the total number $N_p$ of quasi-particles 
\be
N_p(t) = \sum_k \av{\pop{k} (t)}
\ee\label{eq:forplot}
 and the Ising energy
\be
	E_I (t) = \av{H_0 (g,t)} = \suml{k>0}{} \epsilon_k \av{\pop{k} (t)},
\label{eq:E_0}
\ee
which represent two examples of a wider class of quantities which can be generically expressed as
		\be
			\ob (t) = \suml{k>0}{} C_k \av{\pop{k} (t)},
		\label{eq:obclass}
		\ee
		corresponding to $C_k = 1$ and $C_k = \epsilon_k$, respectively.
The evolution of $N_p$ and $E_I$ has been investigated in Ref.~\onlinecite{M-preth}, highlighting the presence of similar prethermal plateaus. The oscillations preceding these plateaus decay algebraically as $t^{-\alpha}$, with $\alpha\simeq 3$. 
%
		The fact that the same exponent $\alpha$ appears in both quantities is likely to be related to the fact that they belong to the same class defined by Eq. \reff{eq:obclass}. Accordingly, we can reasonably expect this exponent to characterize the whole set of observables $\ob (t)$ (see Eq. \eqref{eq:obclass}),  possibly apart from very specific choices of the coefficients $C_k$. 

		In Fig.~\ref{fig:ENcoll} we display $N_p$ and $E_I$ for various  values of $N$ and $\lambda$, showing that the typical amplitude of the oscillations around the corresponding quasi-stationary values ($E_{I,qs}$ and $N_{k,qs}$) scales as $\lambda N$ as long as $\lambda$ is sufficiently small  (i.e., $\lambda \lesssim 0.5$) and $N$ sufficiently large  ($N \gtrsim 40$). Note, however, that the recurrence time $t_R'$ explicitly depends on $N$, thus curves with different $N$ can be collapsed one onto the other, as in Fig.~\ref{fig:ENcoll}, only until the first recurrence appears. On the other hand, this proves that the exponent $\alpha$ of the algebraic decay  depends on neither  $N$ nor $\lambda$. 
%

\end{section}

\begin{section}{Range of validity of the truncation}

\label{sec:validity}

We address here the  range of validity of the truncated Holstein-Primakoff expansion  and its dependence on the quench parameters.
First, we assess the accuracy of the approximation on the basis of the comparison between the bosonic formalism introduced above (see Eq. \eqref{eq:Bogoliubov1}) and the full diagonalization of the fermionic Hamiltonian \reff{eq:chernabog}. Although the latter is still restricted to the totally-paired sector, it represents an exponentially-complex problem, which we could diagonalize only up to $N=20$ sites. 
		\begin{figure}[!ht]
		\centering
		\includegraphics[width =   \columnwidth]{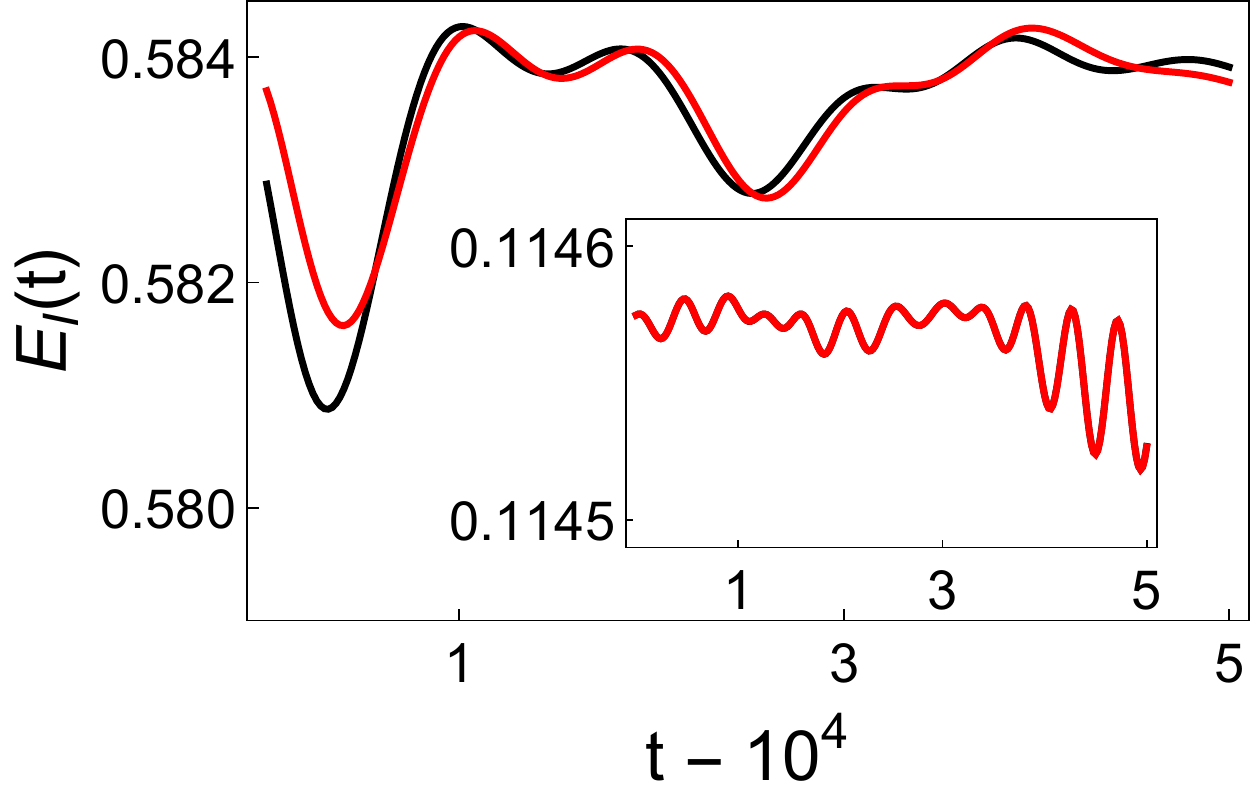}
		
		\caption{(\emph{colour online}) Temporal evolution of the Ising energy $E_I (t)$ (Eq. \eqref{eq:E_0}), after a quench from $g_0 = 8$ to $g=1.5$, $\lambda = 0.1$ and $N= 20$, with $t \simeq 10^4$. Black lines represent data calculated from the exact diagonalization of the original fermionic Hamiltonian \eqref{eq:FH}, whereas red ones refer to the bosonic approximation, Eq. \eqref{eq:freeH'}. This shows that at very long times (see the time scales on the horizontal axis) the latter starts to deviate from the actual evolution, signalling a breakdown of the low-density approximation, upon which it is based. However, this breakdown occurs at increasingly longer times as the post-quench value of $g$ departs from the critical point in the paramagnetic phase, as exemplified by the curves in the inset, which display  the same two quantities but for $g = 3.5$, which are practically indistinguishable.
		}
		\label{fig:compI}
		\end{figure}
Still, this allows us to verify that there is a wide range of values of the quench parameters for which the approximation and the original model are quantitatively almost indistinguishable.

This is demonstrated, e.g., by Fig.~\ref{fig:compI}, where we report the evolution of the Ising energies $E_I^{(b/f)}$ (see Eq.~\eqref{eq:E_0})
		calculated in the bosonic $(b)$ [see Eq. \eqref{eq:freeH'}] and fermionic $(f)$ representation [i.e., the full fermionic Hamiltonian, Eq. \eqref{eq:FH}]  for $N = 20$, $g_0 = 8$, $\lambda = 0.1$ and two different values of $g$ within the time frame $t\in\lqq 10^4, 10^4 + 5 \rqq$. As shown at the end of Sec. \ref{sec:num}, this temporal window  includes time scales much longer than those typically required for pre-thermal features to emerge.
		
		As expected, as the post-quench field $g$ approaches the critical value $g_c = 1$, the approximation becomes less accurate. 
This agrees with the intuitive picture that the closer $g$ is to $g_c$, the larger the initial populations $\av{\pop{k} (t=0)}$ are. 
 As a matter of fact, we have numerically verified that for $g = 1.01$ and all the other parameters fixed as above, the agreement between the two curves in Fig. \ref{fig:compI} remains within $2$\% up to $t \simeq 10^3$, which implies that, as long as the interaction $\lambda$ is small, the low-density approximation features a considerably wide range of applicability. In order to probe the effects of the interaction, we have performed an analogous comparison upon varying $\lambda$. We verified that the relative discrepancy
	\be
			\epsilon(t) = \frac{E_I^{(f)}(t) - E_I^{(b)}(t)}{E_I^{(b)}(t)}
	\label{eq:err}
	\ee
between the kinetic energy computed in the bosonic and fermionic represenatations,  increases upon increasing the interaction $\lambda$. Still, for intra-phase quenches this relative discrepancy remains small also for very long times: for instance, for $g_0 = 8$, $g=3.5$, and $\lambda = 0.9$ we have verified that $|\epsilon|$ keeps within $0.2 \%$ up to the time scales $t\sim 10^4$ probed in Fig.~\ref{fig:compI}:
%
A numerical analysis carried out for smaller system sizes also suggests that the accuracy improves upon increasing $N$. Quenches within the ferromagnetic phase also display low discrepancies for very long times, whereas for quenches across the phases these become important on much shorter time-scales (e.g., for $g_0 = 3.5$, $g = 0.8$ and $\lambda= 0.1$ one finds that $|\epsilon|$ is larger than $1 \%$ already for $t \lesssim 10^3$).


Due to the interacting nature of the model in Eq.~\eqref{eq:FH}, it is not possible to employ the exact diagonalization for a direct comparison much beyond the system size considered in  Fig.~\ref{fig:compI}. However, an indirect measure of the accuracy of  the approximation can also be extracted from the effective integrable bosonic description, Eq.  \eqref{eq:freeH'}. As we prove in Appendix ~\ref{app:HP}, as long as the bosonic occupations $\wh{N}_k\equiv a^\dag_k a_k$ satisfy, in the course of the quench dynamics, the inequality
\be
	\av{\wh{N}_k^2} - \av{\wh{N}_k} \ll \av{\wh{N}_k}
\ee
(where we averaged over the pre-quench state) the  effects of the terms neglected in the truncation can be expected to be small. When this condition is not satisfied, the approximate dynamics departs from  the real one, making the approximation no longer accurate.
%
\begin{figure}[ht]
		\centering
		\parbox{0.9\columnwidth}{
		\subfigure[]{\includegraphics[width = 0.9\columnwidth]{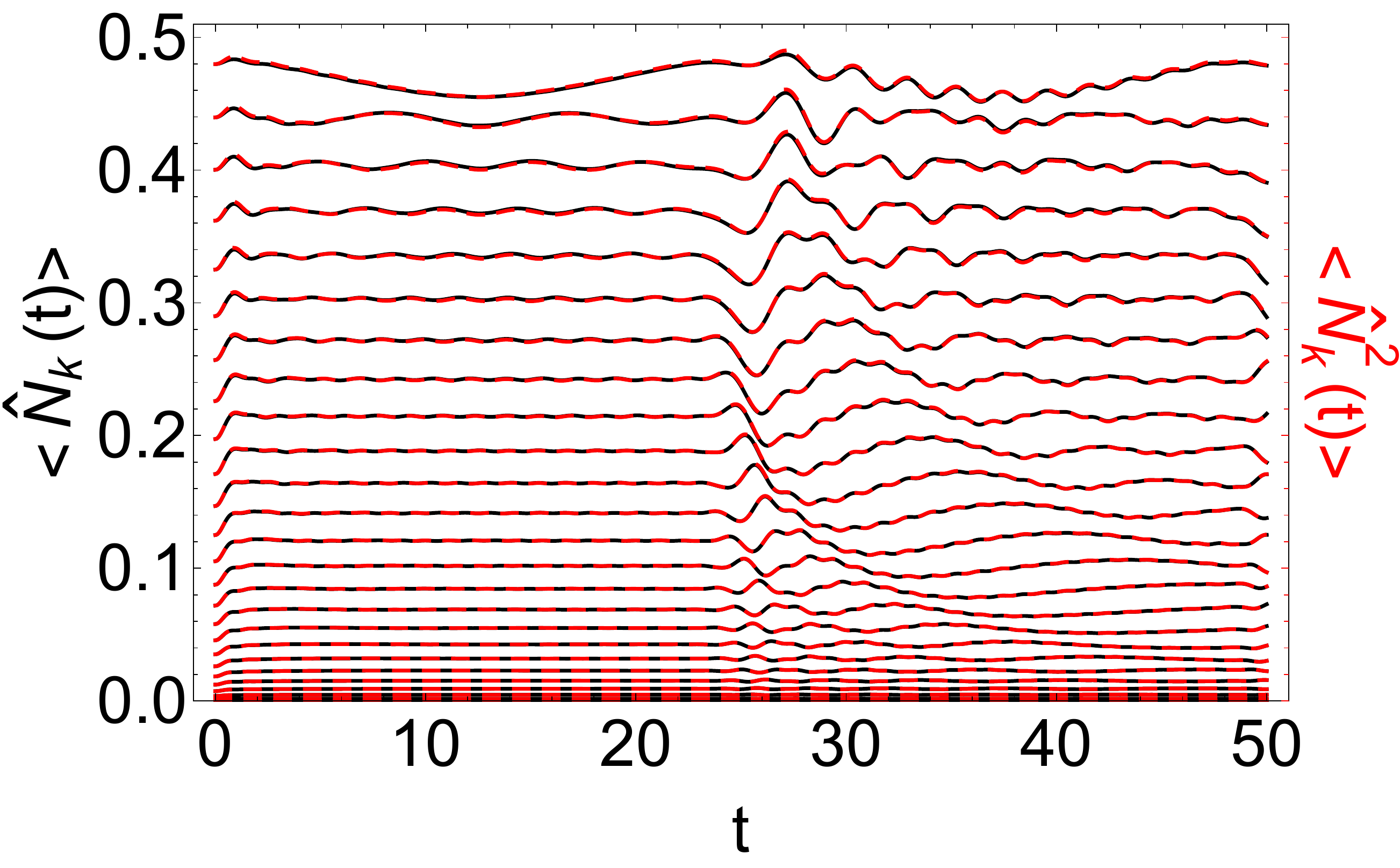} \label{sfig:Nsquare} }}
			\\
		\hspace{-10mm} \parbox{0.85\columnwidth}{
		\subfigure[]{\includegraphics[width = 0.85\columnwidth]{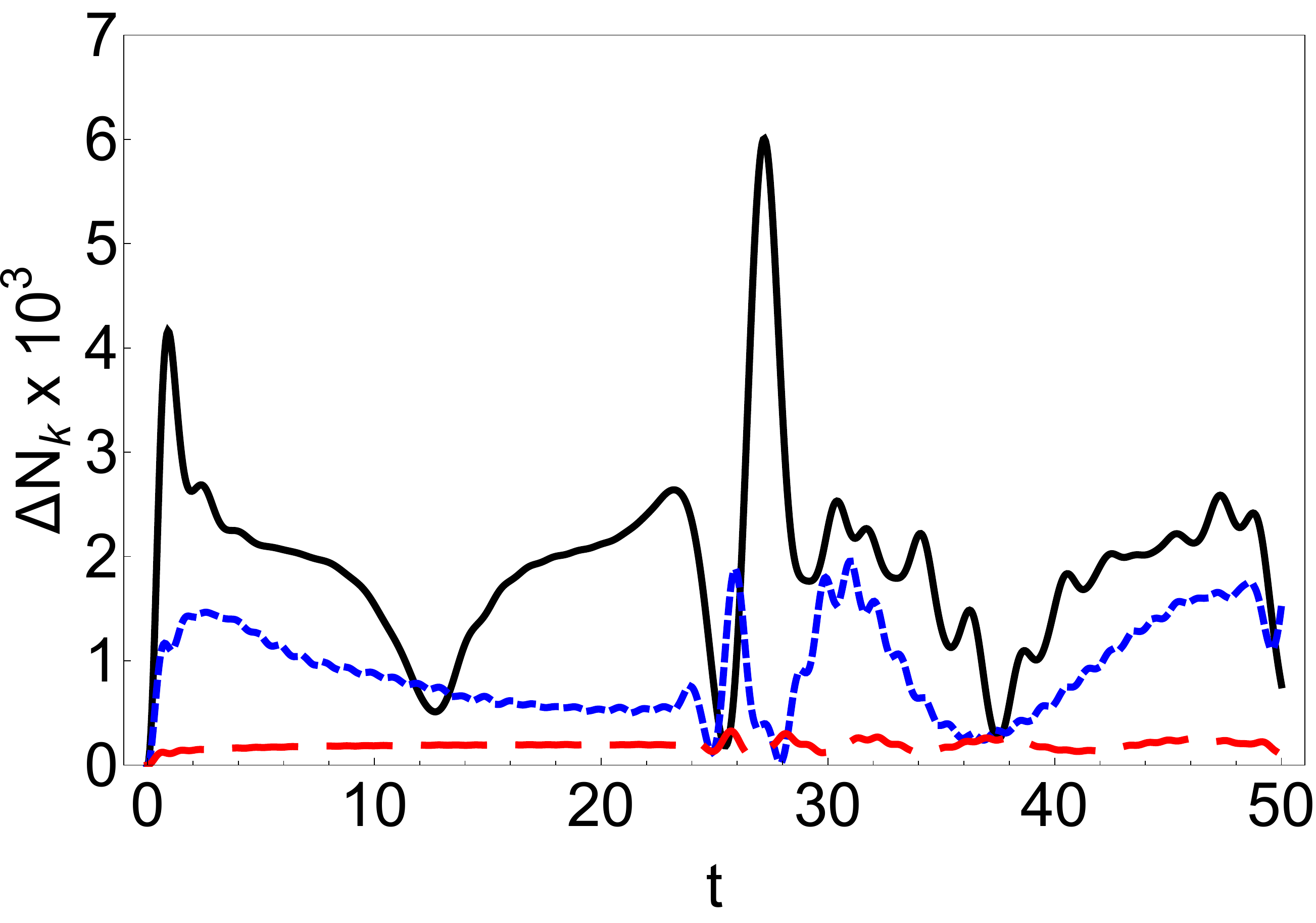} \label{sfig:err2}} } 
		\caption{ (\emph{colour online}) (a) Evolution of the bosonic populations $\av{\wh{N}(t)}$ (black, solid lines) and squared populations $\av{\wh{N}^2(t)}$ (red, dashed lines) for a critical quench from $g_0 = 8$, $g = 1$, with $\lambda = 0.2$ and $N = 50$. From the topmost ($k = \pi/50$) to the lowermost curve ($ k = 49 \pi / 50$), the momentum is monotonic decreasing in steps of $N/2 = 25$ values. This plot reaches $t = 50 \approx 2 t_R$ (with $ t_R \approx N/2 = 25$) and therefore shows that the prethermal regime is still well-described by the free-boson approximation, despite the presence of modes with a large population close to $0.5$. (b)  Evolution of the corresponding relative error \eqref{eq:bnerr} for momenta $k = \pi/50$ (black solid line), $29\pi/50$ (blue short-dashed line, in the middle) and $43\pi / 50$ (red long-dashed line, at the bottom). 
		}
		\label{fig:N^2}
		\end{figure}		
In Fig.~\ref{sfig:Nsquare} we show the evolution of both $\av{\wh{N}_k }$ (left scale, black lines) and $\av{\wh{N}_k ^2}$ (right scale, red lines) for a chain of $N=50$ spins up to time $t = 50$, corresponding to approximately twice the recurrence time $t_R \approx 25$ (the curves in Fig. \ref{fig:N^2} are displayed for various values of wave-vector $k$ from top to bottom). This shows that the two quantities are still closely related within the prethermal regime and that the free-boson approximation is still able to capture the relaxation reasonably well, even for highly-populated modes. Figure \ref{sfig:err2} shows the corresponding relative error
\be
	\Delta N_k \equiv \frac{\abs{\av{\wh{N}_k(t)}  - \av{\wh{N}^2_k(t)}}}{ \av{\wh{N}_k(t)} }
	\label{eq:bnerr}
\ee 
for three different values of the momentum $k$. Interestingly enough, lower-populated modes display smaller errors compared to higher-populated ones. The worst case among the ones displayed in panel (b) corresponds to momentum $k = \pi/50$ within this time window, and it anyhow  shows a discrepancy smaller than $1\%$. This constitutes thereby an upper bound on the error committed by employing this approximation in these parameter ranges. As expected, the maximal error becomes much smaller for intra-phase quenches (within the same time range as the one in Fig. \eqref{sfig:err2}, it remains $< 10^{-6}$ with $g = 3.5$ and $N=30$, data not shown) and increases when crossing the critical point (oscillating around $1 \%$ for $g = 0.5$ and $N = 30$). Furthermore, the relative error increases for quenches across the two phases and it  displays a weaker increase upon increasing  $N$.

\end{section}

\begin{section}{Conclusions}

In this work we reported on a viable approach to study the dynamics of a quenched interacting  quantum spin chain with long-range interactions, which originates from the perturbation of a one-dimensional quantum Ising model. 

We employed a composite quench of the Ising transverse field and of the interaction coupling, in order to initiate the non-equilibrium dynamics. While the quench of the transverse field does not affect  the conserved quantities of the Ising chain (the occupation number of Bogoliubov quasiparticles), the interaction introduces scattering among them, reducing  the integrals of motion of the model to a smaller subset. 

For the specific quench protocol and initial states considered here, the study of  the non-equilibrium dynamics can be based on a lowest-order Holstein-Primakoff expansion in the density of the quasi-particles injected after the quench (see Eq. \eqref{eq:HPb}). Specifically, for quenches that do not cross the critical point, the density of quasi-particles produced is small enough to allow a mapping of the model and of its intermediate-time dynamics into the effective evolution of an integrable model of fully-connected bosons (see Eq. \eqref{eq:chernabog1}). This allows us to extract the algebraic relaxation of a number of observables towards a  state which  constitutes  the asymptotic steady state of the system in the thermodynamics limit: for $N\to\infty$ the model becomes exactly solvable, since the long-range nature of the interaction suppresses fluctuations and makes the mean-field theory exact  (in this respect see Ref.~\onlinecite{fagottibertini}). 

On the other hand, for large but finite system sizes, the steady-state approached by observables is expected to be destabilized in the long-time limit, leading eventually to a thermal steady state, as a result of the energy redistribution due to the scattering among quasi-particles. In Ref.~\onlinecite{M-preth} this aspect has been explored via a kinetic equation  for the time averages of occupation numbers of the Ising quasi-particles. 
In addition to that, we think that inspecting, in the future, the impact of higher order terms in the Holstein-Primakoff expansion could provide insight into the relaxation dynamics of this model; they could be included, for instance, by writing down kinetic equations for two-point functions in the bosonic basis, and numerically studying their evolution after the quench.
This kind of analysis could provide important information on the time scales of the departure from the pre-thermal state, which is a topic of current interest both for theory \cite{Buchhold, Bertini15} and experiments \cite{Langen2016}.
Finally, since the interest towards pre-thermalization in long-range interacting models has increased in recent years, it would be tempting to apply the method presented here to the investigation of those cases \cite{Worm, Morigi2014, Nessi2014, Kastner}.

\end{section}

\acknowledgements

We would like to thank E. Canovi for helpful comments on the numerical diagonalization of the model and P. Calabrese, M. Fabrizio and E. Tonni for useful discussions. J. M. acknowledges support from the Alexander von Humboldt foundation.




\appendix

\begin{section}{The  quantum Ising chain in a transverse field}
	\label{ssec:QIsing}

	The quantum Ising chain in a transverse field consists of a one-dimensional lattice of $N$ sites, each accommodating a  quantum $1/2$-spin. These spins are simultaneously subject to a ferromagnetic nearest-neighbour interaction of strength $J > 0$, which favours configurations in which they are all aligned along a specific spatial direction (say, $x$), and an external magnetic field $Jg$ directed orthogonally to it (e.g., towards $z$), which instead tends to destroy such an ordering. The corresponding Hamiltonian is \cite{QPT}
	\begin{equation}
	\label{eq:IsingH}
		H_0(g)=- \frac{J}{2} \suml{i=1}{N} \left( \hsi_i^x \, \hsi_{i+1}^x + g\hsi_i^z \right).
	\end{equation}
	Periodic boundary conditions are chosen, such that $\sigma_{N+1}^{x/y/z} \equiv \sigma_{1}^{x/y/z}$.
	Here $\hsi^\mu_i$ ($\mu = x, y, , z$) denotes the standard spin operators acting on the $i$-th site. In the eigenbasis of $\hsi^z$ they can be conveniently represented as Pauli matrices:
\be
	\sigma^x = \sxM , \quad \sigma^y = \syM , \quad \sigma^z = \szM.
	\label{eq:Pauli}
\ee
For simplicity, we  set $J = 1$, which is tantamount to  measuring energy and time in units of $J$ and $J^{-1}$, respectively. 
By construction, operators acting on different sites commute:
	\be
		\comm{\hsi^\mu_i}{\hsi^\nu_j} = 0 \quad \quad \ \mbox{with} \quad i \neq j,
	\ee

and for generic $\mu$ and $\nu$.
	The Hamiltonian \reff{eq:IsingH} is  invariant under the $\Z_2$ transformation $\hsi_i^x \to -\hsi_i^x$, $\hsi_i^z \to \hsi_i^z$, which is implemented via the unitary operator $U_{\Z_2} = \prod_i \hsi_i^z$.  This model undergoes a prototypical quantum phase transition\cite{QPT} at the critical value $g = g_c = 1$; for $g>g_c$ the system is {paramagnetic} and the {longitudinal magnetization} $\av{\hsi_i^x} $ identically vanishes, whereas for $g<g_c$ a {ferromagnetic} ordering ensues which entails a spontaneous breaking of the $\Z_2$ symmetry, i.e., $\av{\hsi_i^x} \neq 0$.

The inherent integrability of the Hamiltonian in Eq.~\reff{eq:IsingH} is made apparent after a Jordan-Wigner transformation accompanied by a Bogoliubov rotation\cite{QPT}. The former reads\cite{JW, JW2}
	\be
		\sysb{l}
			\hsi_i^x = \prodl{j=1}{i-1} \lt 1 - 2c_i^\dag c_i \rt \lt c_i^\dag + c_i \rt , \\[2mm]
			\hsi_i^y = i\,\prodl{j=1}{i-1} \lt 1 - 2c_i^\dag c_i \rt \lt c_i^\dag - c_i \rt ,  \\[2mm]
			\hsi_i^z = 1 - 2c_i^\dag c_i,
		\syse
	\ee
	and is expressed in terms of fermionic creation and annihilation operators,  $c_i^\dag$ and $c_i$ respectively, with
\be	
	\acomm{c_i}{c_j^\dag} = \delta_{ij} \quad \mbox{and} \quad \acomm{c_i}{c_j} = 0.
\ee
	The Hamiltonian in this new basis reads
	\be
	\begin{split}
		H_0(g) = -\frac{1}{2} \suml{i=1}{N-1} \lqq  c_i^\dagger c_{i+1} + c_{i+1}^\dagger c_{i} +  c_{i}^\dagger c_{i+1}^\dagger + c_{i+1} c_i  \rqq + \\
		+ \frac{1}{2} U_{\Z_2} \lt c_N^\dagger c_{1} + c_{1}^\dagger c_{N} +  c_{N}^\dagger c_{1}^\dagger + c_{1} c_N  \rt - g \lt \frac{ N}{2} -  \wh{N} \rt,
	\end{split}
	\label{eq:IH1}
	\ee
with $\wh{N} = \sum_i c_i^\dag c_i$, and it is translationally-invariant (with anti-periodic and periodic boundary conditions in the even- and odd-parity sectors, respectively) and quadratic. It can be therefore mapped to a free-fermion model via a Bogoliubov rotation\cite{QPT} (including a Fourier transform)
\be
\gamma_k=u_kc_k-iv_kc_{-k}^\dag
\ee
with the new operator basis   denoted by $\gamma_k$, $\gamma_k^\dag$, and $k$ the quasi-momentum running over discrete values
\be 
	k_n = \frac{2\pi}{N} \lt n-\ha  \rt \quad \mbox{with} \quad n = -\frac{N}{2} + 1, \ldots , \frac{N}{2}.
	\label{eq:kval}
\ee 
The coefficients $u_k=\cos\theta_k(g)$ and $v_k=\sin\theta_k(g)$ are defined through 
 the Bogoliubov angle   $\theta_k(g)$ given by \cite{QPT}
	\be
		\tan (2\theta_k(g)) = \frac{\sin k}{g - \cos k}.
	\label{eq:bangle0}
	\ee

Therefore, also the operators $\gamma_k(g)$ and $\gamma^\dag_k(g)$, depend on $g$ and  they keep   preserved the canonical anticommutation relations
	\be
		\acomm{\gamma_k}{\gamma_q^\dag} = \delta_{kq}, \quad \acomm{\gamma_k}{\gamma_q} = \acomm{\gamma_k^\dag}{\gamma_q^\dag} = 0.
	\label{eq:canacomm}
	\ee	
We employ here the same convention as that one introduced in the main text, i.e., all quantities refer to the post-quench value $g$ unless otherwise specified.
An additional $k \leftrightarrow -k$ symmetry of the Hamiltonian makes it possible to express it in the compact form
%
%
%
%
%
%
	\be
		H_0 (g) =  \suml{k > 0}{} \epsilon_k \psi_k^\dag \sigma^z \psi_k\,,
	\label{eq:IsingH1}
	\ee
	where the summation is restricted to the positive values of $k$ in Eq.~\eqref{eq:kval} and
	\be
		\psi_k = \matb{l} \gamma_k\\ \gamma^\dag_{-k} \mate \quad \text{ and } \quad \psi^\dag_k = \matb{l} \gamma^\dag_k\\ \gamma_{-k} \mate^\intercal
		\label{eq:spinor}
	\ee
	are Nambu spinors (here $\intercal$ denotes transposition),
	while	
	\be
		\epsilon_k\equiv\sqrt{1+g^2  -  2 g\cos k }
	\label{eq:disprel}
	\ee
is the dispersion relation of the quasi-particles. 
%
%
%
%
%
%

Accordingly, the Hamiltonian \reff{eq:IsingH1} is non-interacting: the various modes are independent and preserved under the unitary evolution, i.e., all the fermionic \emph{populations}
	\be
		\hat{n}_k = \gamma_k^\dag \gamma_k
	\label{eq:pop1}
	\ee
commute with $H_0(g)$ and therefore they do not evolve. These operators constitute a sufficiently large set of conserved quantities to analytically solve the model. The simplicity  of this solution might appear to be conflicting with the complexity of the collective modes driving a phase transition; however, the Jordan-Wigner transformation connecting the free-fermionic and the spin models is both non-linear and non-local and makes the correlation function of the order parameter a highly non-trivial combination of the free-fermion expectation values.

The $N$ constraints introduced by the conservation of the populations \reff{eq:pop1}  are actually much stringent than the original $\Z_2$ symmetry mentioned above, as the latter can be implemented in this picture by the operator
	\be
		U_{\Z_2} = \prodl{k}{} (1 - 2 \hat{n}_k) = \rme{i\pi \suml{k}{} \hat{n}_k},
	\label{eq:Z2-1}
	\ee
which describes the parity of the total number $N_p$ of fermions. In other words, $U_{\Z_2}$ evaluates to $1$ if $N_p$ is even and to $-1$ if odd. Note that the rightmost equality above comes from the fact that $\hat{n}_k^m \equiv \hat{n}_k $ for every integer $m \ge 1$, due to the anticommutation relations in Eq. \reff{eq:canacomm}.

\end{section}

\begin{section}{Subtraction of $\overline{M}^z$}
	\label{subtrcapp}

The interaction term $V$ in Eq.~\reff{interaction}, which is added to the Hamiltonian of the Ising chain, is long-range in nature since the total transverse magnetization $M_z$ is an extensive quantity, requiring such a term to be explicitly divided by the system size $N$ in order to guarantee the extensivity of the energy. $\overline{M}^z$ in Eq. (2) indicates the corresponding long-time average calculated for $\lambda = 0$, i.e., according to Eq. \eqref{eq:Mz_bar}.
	The subtraction in Eq. \eqref{interaction} is meant to cancel the ``integrable'' part of the operator $M^z$, i.e., the constants of motion $\hat{n}_k$ which enter its definition (see Eqs.~\reff{eq:Mz1} and \reff{eq:Mz_av} below)  and, indeed, one can prove that the connected correlation function of the remainder $\Delta M^z\equiv M^z-\overline{M}^z$ satisfies, in the thermodynamic limit, the cluster property at long times
	\be
	\label{corrsottratto}
		\liml{\tau \to \infty} \,\liml{t \to \infty} \,\lan  \Delta M^z(t+\tau)    \, \Delta M^z(t)  \ran_c = 0,
	\ee
with $\Delta M^z = M^z - \overline{M}^z$. We now analyze more closely Eq. \eqref{corrsottratto}. We first consider the evolution of $M^z$ in the Heisenberg picture,
\be
\begin{split}
	M^z(t) & =  \rme{i \, H_0 (g)\, t}\,  M^z  \, \rme{-i \, H_0 (g)\, t} =  \\
	& =2\sum_{k>0}\cos (2\theta_k(g))(\gamma_k^\dag   \gamma_k-\gamma_{-k}\gamma_{-k}^\dag) + \\
	&-2i\sum_{k>0}(\sin 2\theta_k(g))({\rm e}^{2i\epsilon_k t}\gamma_k^\dag\gamma_{-k}^\dag-{\rm e}^{-2i\epsilon_k t}\gamma_{-k}\gamma_{k}),
\label{eq:Mz1}
\end{split}
\ee
where $\theta_k(g)$ is the Bogoliubov angle. 
The time average in Eq. \eqref{eq:Mz_bar} wipes out the oscillating factors, yielding
\be
\overline{M^z}= 2\sum_{k>0}\cos (2\theta_k(g))(\gamma_k^\dag\gamma_k-\gamma_{-k}\gamma_{-k}^\dag).
\label{eq:Mz_av}
\ee
and, correspondingly,
\be
\Delta M^z  = -2i\sum_{k>0}\sin (2\theta_k(g))(\gamma_k^\dag\gamma_{-k}^\dag-  \gamma_{-k}\gamma_{k}).
\label{eq:dMz}
\ee
On the other hand, the long-time limit of the the two-times correlation function --- without subtraction of $\overline{M}_z$, is a non-vanishing constant $C$ (see, for instance Refs. \onlinecite{Foini2} and \onlinecite{Mazur}), 
\be
		\liml{\tau \to \infty} \,\liml{t \to \infty} \,\lan  M^z(t+\tau)    M^z(t)    \ran =C\neq0,
	\ee
which is equal also to $C=\langle (\overline{M}^z)^2\rangle-\langle \overline{M}^z\rangle^2$.  The correlation function of $M^z$ after a quantum quench,  spoils the time clustering property, because in the initial state there are non-vanishing correlations $\langle \gamma^\dag_k(g)\gamma^\dag_{-k}(g)\rangle\neq0$ and $\langle \gamma_{-k}(g)\gamma_{k}(g)\rangle\neq0$ between modes of opposite momenta. Remarkably, since the global nature of the operator $M^z$ implies that $C$ scales with the system size, i.e., $C\sim O(L)$, the long-time clustering property is spoiled in the thermodynamic limit $L\to\infty$. 

Notice, for comparison, that $C$ is also present in correlation functions of local operators, like the on-site transverse magnetization correlation function $\lan\sigma_{i+r}^z\sigma_i^z\ran$ but there the contribution by $C$ turns out to be  subleading in the thermodynamic limit, i.e.,  $\sim O(L^{-1})$. Accordingly, the  subtraction of $\overline{M}^z$, in the interaction term $V$ removes a sort of undesired memory-preserving behaviour.

\end{section}

\begin{section}{The Holstein-Primakoff transformation and its truncation}

\label{app:HP}

Hard-core bosons emerge naturally in the context of  quantum spin-$1/2$ chains: it is in fact sufficient to introduce the spin ladder operators $\hat{\sigma}^{\pm}_i$ as
		\be
			\hsi_i^\pm = \frac{1}{2} \lt \sigma_i^x \pm \sigma_i^y  \rt,
		\ee
		where $\hsi_i^{\mu}$ are the spin operators already encountered at the beginning of Appendix \ref{ssec:QIsing}, to yield the identification
		\be
			\hsi_i^+ \,\leftrightarrow\, b_i^\dag   ,    \quad \quad \hsi_i^- \,\leftrightarrow\, b_i   ,     \quad \quad \hsi_i^z \,\leftrightarrow\,  2b_i^\dag b_i  -1,
		\ee
where indeed $b_i^\dag$ and $b_i$ act as creation and annihilation operators of hard-core bosons. The Hilbert space can be accordingly reinterpreted by setting the correspondence $\ket{\uparrow_i} \,\leftrightarrow \, \ket{{\bf 1}} $, $\ket{\downarrow_i} \,\leftrightarrow \, \ket{{\bf 0}}$, in analogy with the notation introduced after Eq.~\eqref{eq:almostcomm}: $\ket{{\bf 0}}$ denotes absence of the hard-core boson, whereas $\ket{{\bf 1}}$ the presence of a single one.
The hard-core constraint can be reinterpreted as emerging from strongly  repulsive contact interactions among the quasi-particles. 

The Holstein-Primakoff transformation\cite{HP}
expresses $b_i$ and $b^\dag_i$ in terms of non-linear functions of the standard (free) bosonic operators $a_i$ and $a_i^\dag$. Here we shall focus, for simplicity, on a single mode $b$, $b^\dag$, (by dropping the site index) which we recast in the form
		\be
			b = \sqrt{\id - \wh{N}} \,\, a,   \quad \quad b^\dag = a^\dag \sqrt{\id - \wh{N}},
		\label{eq:HP2}
		\ee
		with $a$ and $a^\dag$ obeying the usual commutation relations $[a,a^\dag] = \id$ and $\wh{N} = a^\dag a$. The Hilbert space is enlarged accordingly, from the two-dimensional space spanned by $\ket{{\bf 0}}$ and $\ket{{\bf 1}}$, to the infinite-dimensional one generated by the usual bosonic number basis $\left\{ \ket{n} \right\}_{n \in \N}$. On the other hand, the latter is split into two sectors which cannot be connected by $b$ and $b^\dag$, and which respectively include all the ``physical'' states $\left\{ \ket{n} \right\}_{n = 0,1}$ and all the ``unphysical'' ones $\left\{ \ket{n} \right\}_{n >1}$. This represents a relevant aspect, as the anticommutation relations
		\be
			\acomm{b}{b^\dag} =  \id + 2\wh{N} (\id - \wh{N})
		\ee
		are not correctly reproduced at the operatorial level; however, on both physical states (and, thus, in the whole physical subspace) one finds $\acomm{b}{b^\dag}  \equiv \id$,
		thereby recovering the hard-core nature. When expanding the square roots in Eq.~\reff{eq:HP2} as power series, and approximating the $b$ operators by truncation at any finite order, the separation between the physical and unphysical subspaces becomes weaker. This fact emerges quite clearly when considering the simplest possible case, i.e., $b^{(\dag)} \approx  a^{(\dag)}$; indeed, within this approximation, we find in fact that $b^\dag \ket{1} = \ket{2}$, which connects the physical state $\ket{1}$ with the unphysical one $\ket{2}$. Consequently, the regime of validity of such an approximation is determined by the overlap of the state under study with the physical basis: the more it resembles its projection onto the physical space, the more accurate the result is. 

We shall now briefly discuss the implications that the truncation has on the populations $\pop{} = b^\dag b$. 
		In Sec.~\ref{sec:num}, we extensively used the approximation $\pop{} \approx \wh{N}$, which is valid only in the physical sector: in fact, if we restrict the states to just $\ket{0}$ and $\ket{1}$, we can see that
		\be
			\whN^m \ket{0} = 0 \quad \text{ and } \quad \whN^m \ket{1} = \ket{1}
		\label{eq:whN}
		\ee
		for every (integer) $m\neq0$. Thus, as long as the system is still lying approximately in the physical space, we can approximate $\whN^m \simeq \whN$, which renders $\pop{} \simeq \whN$. Conversely, it is true that if $\av{\whN^m} \simeq \av{\whN}$, i.e.,
		\be
			\frac{\av{\whN^m} - \av{\whN}}{\av{\whN}} \ll 1,
		\label{eq:low-dens1}
		\ee
	for any integer $m \geq 2$, 	then the truncation holds, as we prove below. 
		
		Consider, in fact, a generic normalised state $\ket{\psi} = \sum_n a_n \ket{n}$; according to the discussion above, $\ket{\psi} $ can be considered ``approximately physical'' as long as $\sum_{n>1} \abs{a_n}^2 \ll 1$. Calculating the averages in Eq.~\reff{eq:low-dens1} on this state one can rewrite this condition  as
			\be	
			\suml{n = 2}{\infty} \abs{a_n}^2 \lt n^m -  n \rt \ll \abs{a_1}^2 + \suml{n = 2}{\infty} \abs{a_n}^2 n;
		\label{eq:low-dens3}
		\ee
		in turn this equation	is equivalent to
		\be
			\suml{n = 2}{\infty} \abs{a_n}^2 \lt n^m -  2n \rt \ll \abs{a_1}^2.
		\ee
		Using the equation above, it is easy to prove that the state, $|\psi\rangle$, which satisfies \eqref{eq:low-dens1} is  indeed an ``almost physical'' one. 

		From this discussion, we understand that, although the validity of the truncation may be intuitively expected to rely on a small quasi-particle density  as suggested by the name ``low-density approximation'', it relies instead on higher moments of the quasi-particle densities, as shown by the condition  \eqref{eq:low-dens1}. On the other hand, $\av{\whN} > 1$ constitutes a clear sign that unphysical states are populated; therefore, it is still physically meaningful to focus on a low-density condition $\av{\whN} \ll 1$.

%

\end{section}

	
	\begin{section}{Diagonalization of the bosonic Hamiltonian (Eq. \eqref{eq:freeH'}) and calculation of the observables}
	\label{app:numdiag}
	
		The Bogoliubov rotation in Eq.~\reff{eq:Bogoliubov1}, which diagonalises $H'$, constitutes a change of basis and can therefore be expressed by a suitable unitary transformation. On the other hand, its implementation \reff{eq:Bogoliubov1} is not realised via a unitary matrix. As a matter of fact, for the commutation relations to be preserved, the $(N/2) \times (N/2)$ matrices $A$ and $B$, constructed from $A_{kq}$ and $B_{kq}$, respectively, in Eq. \eqref{eq:Bogoliubov1}, must obey the identities
		\be
			A A^\dag - B B^\dag = \id   ,   \quad \quad A B^\intercal - B A^\intercal = 0,	
		\label{eq:prescomm}	
		\ee
		which define a symplectic matrix (see Appendix ~\ref{sapp:Williamson})
		\be
			M = \matb{c|c} A & B \\ \hline B^\ast & A^\ast    \mate,
		\label{eq:NxN}
		\ee
		
		acting as $M \vec{\eta} = \vec{a}$ on the vectors
		
		\be
		\begin{split}
			\vec{\eta} & = \lt \eta_1, \eta_2 , \ldots , \eta_{N/2}, \eta_1^\dag, \eta_2^\dag ,  \ldots , \eta_{N/2}^\dag  \rt, \\
			\vec{a} & = \lt a_1, a_2 , \ldots , a_{N/2}, a_1^\dag, a_2^\dag ,  \ldots , a_{N/2}^\dag  \rt,
		\end{split}
		\ee
		where $\vec{\eta}$ are the Bogoliubov operators (see Eq. \eqref{eq:Bogoliubov1}) which diagonalize the bosonic Hamiltonian (see Eq. \eqref{eq:chernabog1}) defined through the bosonic operators $\vec{a}$.
		The diagonalization procedure reported in Appendix ~\ref{sapp:Williamson}, which makes use of the constructive proof of Williamson's theorem\cite{Williamson}, is exact in the sense that no approximation is involved other than the low-density one already employed to write down $H'$ in Eq.~\reff{eq:chernabog1}. Single-particle eigenvalues and eigenvectors can therefore be obtained in principle to any desired accuracy. However, this is still not sufficient for studying the dynamics, which in general requires also to rewrite the initial state $\ket{0}$ in terms of the new Fock basis 
		corresponding to the operators $\eta_k^\dag$, $\eta_k$. Note that even though in the $a$-operators basis the state is a combination involving, for each mode $k$, only  states with  $0$ or $1$ particles, this is not generally true in the new basis. This implies the necessity to approximate it with its projection on a finite subspace, thereby spoiling the exactness of the diagonalization. On the other hand, since the system is free, this obstacle can be conveniently overcome by using the Heisenberg picture for the evolution, instead of the Schr\"odinger one: we consider for instance a population $\pop{k}$ and introduce the inverse transformation 
		\be
			M^{-1} = \matb{c|c} C & D \\ \hline D^\ast & C^\ast    \mate,
		\label{eq:anti-Bogoliubov1}
		\ee
		which maps matrices written in the $\eta$, $\eta^\dag$ Bogoliubov basis  back to the bosonic one written in terms of $a$, $a^\dag$: $M^{-1} \vec{a} = \vec{\eta}$. Note that, since the quench dynamics discussed in the main text occurs in a sector with $I_k = 0$, we are allowed to treat the operators $\pop{ k}$ and $\pop{-k}$ as if they were equivalent ($\pop{k} = \pop{-k}$); thus, making use of the identity \reff{eq:opquart}, we can rewrite
		\be
			\pop{k} (t) \eq \pop{k}^2 (t) \eq    \pop{k} (t) \pop{-k} (t)  \eq     b^\dag_k (t) b_k (t) \, \approx \, a^\dag_k(t) a_k (t),
		\ee
		for any time $t$ after the quench, as long as the last approximation holds (we recall that this is the case at $t=0$). Using the change of basis \reff{eq:Bogoliubov1} we find
\begin{widetext}
		\be
		\begin{split}
			\langle a_k^\dag (t) a_k(t) \rangle = \suml{q_1>0, q_2>0}{} \Big\{ 2 \operatorname{Re} \lqq A_{kq_1}^\ast B_{kq_2} \av{\eta_{q_1}^\dag(t) \eta_{q_2}^\dag (t)} \rqq    +      ( A_{k q_1}^\ast A_{k q_2} + B_{k q_1} B_{k q_2}^\ast  ) \av{\eta_{q_1}^\dag(t) \eta_{q_2} (t)} \Big\}   .
		\end{split}
		\label{eq:numpop1}
		\ee
\end{widetext}
		We now make use of the fact that the system is quadratic for explicitly determining the temporal evolution of the operators $\eta_q$ and $\eta^\dag_q$: according to Eq.~\reff{eq:freeH'} we have
		\be
			\eta_q^\dag (t) = \rme{iE_q t} \eta_q^\dag \quad \text{ and } \quad   \eta_q (t) = \rme{-iE_q t} \eta_q,
		\label{eq:evoldiag1}
		\ee
		where $E_q$ are the energies of the Bogoliubov modes determined numerically (see Eq. \eqref{eq:freeH'}).
		Consequently, the expectations in Eq.~\reff{eq:numpop1} oscillate as
		\be
		\begin{split}
			\langle\eta_{q_1}^\dag(t) \eta_{q_2}^\dag (t)\rangle =  \lt Z_1^\dag \rt_{q_1 q_2} \rme{i\lt E_{q_1} + E_{q_2} \rt t}    ,    \\
			\langle \eta_{q_1}^\dag(t) \eta_{q_2} (t) \rangle =  \lt Z_0 \rt_{q_1 q_2} \rme{i\lt E_{q_1} - E_{q_2} \rt t},
		\end{split}
		\label{eq:fastslow}
		\ee
		where the matrices $Z_{0,q_1 q_2} = \langle \eta_{q_1}^\dag \eta_{q_2}\rangle$ and $Z_{1,q_1 q_2} = \av{\eta_{q_1} \eta_{q_2}}$ contain the information about the initial ($t=0$) expectation values of all spin  operators that can be written quadratically in terms of bosons $a_q$ (or equivalently $\eta_q$). 
We introduce the corresponding matrices
		\be
		\begin{split}
			\lt W_0 \rt_{k_1 k_2} &= \langle a_{k_1}^\dag a_{k_2} \rangle = \delta_{k_1 k_2} \sinna{2}{\Delta \theta_{k_1}} +  \\
			&+\lt \frac{1 - \delta_{k_1 k_2}}{4} \rt \sina{\Delta \theta_{k_1}} \sina{\Delta \theta_{k_2}}    ,    \\[2mm]
			\lt W_1 \rt_{k_1 k_2} &= \av{a_{k_1} a_{k_2}} =  \lt \frac{ \delta_{k_1 k_2} - 1}{4} \rt \sina{\Delta \theta_{k_1}} \sina{\Delta \theta_{k_2}},
		\end{split}
		\label{eq:explM1}
		\ee
		where $\Delta \theta_k = \theta_k (g) - \theta_k (g_0)$ and $\theta_k(g)$ is determined according to Eq. \eqref{eq:bangle0}.
		In the $N \times N$ block representation introduced in Eq.~\reff{eq:NxN} these matrices can be reorganised as
		\be
		\begin{split}
			\mal{Z} = \av{\vec{\eta} \otimes \vec{\eta}} =  \matb{c|c} Z_1 & \id + Z_0^\intercal \\ \hline Z_0 & Z_1^\dag    \mate,   \\
			  \mal{W} = \av{\vec{a} \otimes \vec{a}} = \matb{c|c} W_1 & \id + W_0 \\ \hline W_0 & W_1    \mate,
		\end{split}		
		\ee
		where we used the properties $W_0 = W_0^\intercal$ and $W_1 = W_1^\dag$ which can be easily inferred from their explicit forms \reff{eq:explM1}.
		Exploiting the inverse change of basis in Eq. \reff{eq:anti-Bogoliubov1} we finally find
		\be
			\mal{Z} = \av{ M^{-1}\vec{a} \otimes  M^{-1}\vec{a}} = M^{-1} \mal{W} \lt M^{-1}\rt^\intercal,
		\ee
		which allows an exact numerical calculation of the populations, in the sense described above. Unfortunately, this construction, which relies on the Heisenberg picture, has the disadvantage of being specific to the operator chosen; for example, for a quartic one in the operators $a$, it would be necessary to calculate every possible entry of the average $\av{\vec{a} \otimes \vec{a} \otimes \vec{a} \otimes \vec{a}}$, which denotes a $4$-tensor of dimension $(N/2)^4$. On the other hand, once a specific tensor
		\be
			\mal{W}^{(m)} \equiv \langle  \underbrace{ \vec{a} \otimes \vec{a} \otimes \ldots \otimes \vec{a} }_{m \text{ times}} \rangle
		\ee
		has been obtained, the corresponding dynamical expectation
		\be
		\begin{split}
			C_{k_1 \ldots k_m} & (t_1, \ldots t_m) \equiv  \\
			&\lan  \lt\vec{a} \rt_{k_1} (t_1) \otimes\lt \vec{a} \rt_{k_2} (t_2) \otimes \ldots \otimes \lt\vec{a} \rt_{k_m} (t_m) \ran
		\end{split}		
		\ee
		can be in principle calculated for any choice of the times by applying the formula
		\be
		\begin{split}
			C_{k_1 \ldots k_m} (t_1, \ldots t_m) = \lt M \mal{U} (t_1) M^{-1} \rt_{k_1 k_1'}  \times \ldots  \times \\
			\times \lt M \mal{U} (t_m) M^{-1} \rt_{k_m k_m'} \mal{W}^{(m)}_{k_1' \ldots k_m'}.
		\label{eq:nformula}
		\end{split}		
		\ee
		Here $\mal{U} (t)$ keeps track of the evolution in the diagonal basis $\eta$, $\eta^\dag$ and can be written as
		\be
			\mal{U} (t) = \rme{-2i\mal{E}t}, \quad \text{ with } \quad \mal{E} = \quadblock{\bE}{0}{0}{-\bE}
		\label{eq:Ut}
		\ee
		and $\bE$ is the diagonal matrix defined by $2\bE = \operatorname{diag} \left\{ E_1 , E_2, \ldots E_{N/2} \right\}$, where $\left\{ E_i \right\}_i$ is the bosonic one-particle spectrum. We emphasize that, for any  choice of the time coordinates, the only operation left is the contraction of the indices $k_i'$ in Eq.~\reff{eq:nformula}, which just involves $(N/2)^m$ sums. Despite being of polynomial complexity, this can still pose serious difficulties for the investigation of large systems: for instance, the calculation of $\av{\wh{N}_k^2}$ for Fig.~\ref{sfig:Nsquare} is limited to $N=50$ by the fact that, differently from the simpler populations $\av{\pop{k}}$, it involves a $4$-index tensor.


		\begin{subsection}{Williamson's theorem}
		\label{sapp:Williamson}

	  A symmetric, positive-definite, $2n \times 2n$ matrix  can be always brought into diagonal form by a symplectic transformation and  the corresponding spectrum is positive and doubly-degenerate. This is the statement of Williamson's theorem 	\cite{Williamson}. The proof is constructive and shows how to translate the problem into one of standard diagonalization; since the algorithm we have employed follows its main steps \cite{Williamson}, we will report it here. 
	  
	  We start by recalling that a $2n \times 2n$ matrix $S$ is said to be symplectic ($S \in Sp\lt 2n, \R \rt$) if
		\be
			S \Omega S^\intercal = \Omega \quad \text{ with } \quad \Omega = -\Omega^\intercal = \quadblock{0}{\id_n}{-\id_n}{0},
		\label{eq:sympl}
		\ee
		where $\id_n$ is the $n \times n$ identity. As we have mentioned at the beginning of Appendix ~\ref{app:numdiag}, the Bogoliubov rotation \reff{eq:NxN} defines in general a complex symplectic matrix, whereas here we assumed that  it is real. In order to circumvent this complication, we employ the unitary transformation
		\be
			\vec{r} = U \vec{a}   ,   \quad \quad \vec{\rho} = U \vec{\eta} , 
		\ee
		with
		\be
			U = \frac{1}{\sqrt{2}} \quadblock{\id_{N/2}}{\id_{N/2}}{-i \id_{N/2}}{i \id_{N/2}},
		\ee
		which represents (apart from a multiplicative factor) a transformation from ladder operators $a_k$, $a_k^\dag$ to ``position'' and ``momentum'' operators
		\be
		\begin{split}
			r_k = \frac{a_k^\dag + a_k}{\sqrt{2}}, \quad p_k = \frac{i (a_k^\dag - a_k)}{\sqrt{2}},    \\[2mm]
			\rho_k = \frac{\eta_k^\dag + \eta_k}{\sqrt{2}}, \quad \pi_k = \frac{i (\eta_k^\dag - \eta_k)}{\sqrt{2}} .
		\end{split}
		\ee
		Clearly, these new operators obey the canonical bosonic commutation relations $\comm{r_k}{p_q} = \comm{\rho_k}{\pi_q} = i \delta_{kq}$. As a shorthand, we group them in $N$-component vectors according to the basis: $ \vec{r} = (r_1,\ldots,r_{N/2},p_1,\ldots,p_{N/2})$ (starting basis) and $\vec{\rho} = (\rho_1,\ldots,\rho_{N/2},\pi_1,\ldots,\pi_{N/2})$ (diagonal basis). These two vectors are therefore mapped one onto the other via the change of basis
		\be
		\begin{split}
			S = UMU^\dag =  \quadblock{\operatorname{Re}\lt A + B \rt}{\operatorname{Im}\lt  B- A \rt}{\operatorname{Im}\lt A + B \rt}{\operatorname{Re}\lt A - B \rt}
		\end{split}
		\label{eq:NxN1}
		\ee
		which is now evidently a real matrix. To see that $M$ is symplectic we simply apply the definition \reff{eq:sympl} to its form \reff{eq:NxN}, which yields
		\be
		\begin{split}
			M\Omega M^\intercal &  =  \quadblock{AB^\intercal - BA^\intercal}{AA^\dag - BB^\dag}{-\lt AA^\dag - BB^\dag \rt^\intercal}{\lt AB^\intercal - BA^\intercal \rt^\dag}  \\
			& = \quadblock{0}{\id_n}{-\id_n}{0} \equiv \Omega,
		\end{split}
		\ee		
		where on the second line we have used the identities \reff{eq:prescomm}. This implies that the matrix $S$ preserves $\Omega' = U \Omega U^\intercal = i\Omega$, which is tantamount to say that it is symplectic as well. We now proceed to show how this matrix can be determined. Using the canonical commutation relations, we rewrite the Hamiltonian $H'$ in Eq.~\reff{eq:chernabog1} as
		\be
			H' = \vec{a}^\dag \Xi' \vec{a} - \sum_{k_1,k_2} \delta_{k_1 k_2} \beta_{k_1 k_2} \quad \text{ with } \quad \Xi' = \quadblock{\beta}{-\alpha}{-\alpha}{\beta}, 
		\label{eq:cb2}
		\ee
		where $\alpha$ and $\beta$ are the $n \times n$ matrices defined in Eq. \reff{eq:bmatrices} with $n = N/2$ (we recall that we have assumed $N$ to be even). The corresponding form in coordinate space $(r,p)$ is
		\be
			\Xi = U \Xi' U^\dag = \quadblock{\beta - \alpha}{0}{0}{\beta + \alpha}.
		\ee
		Note that $\lt \beta - \alpha \rt_{k_1 k_2} = \delta_{k_1 k_2} \epsilon_{k_1}$ (with $\epsilon_{k}$ given in Eq. \eqref{eq:disprel0}), so that half of this matrix is diagonal and displays the unperturbed eigenvalues $\epsilon_k \geq \abs{g-1} $, which makes it positive definite (if not at the critical point). The other half is given by
		\be
			(\beta + \alpha)_{k_1 k_2} \eq \epsilon_{k_1} \delta_{k_1 k_2} + \frac{2\lambda}{N} \lt 1 - \delta_{k_1 k_2} \rt \sina{2\theta_{k_1}} \sina{2\theta_{k_2}}
		\ee
		and we can safely assume that, as long as $g$ is kept far from $g_c = 1$ and $\lambda$ is not too large, also this part is positive definite  and therefore $\Xi$ satisfies all the requirements of the theorem. This implies that both the inverse $\Xi^{-1}$ and its ``square root'' $\Xi^{-1/2}$ exist and are symmetric, positive-definite matrices. We now define $K = \Xi^{-1/2} \Omega \Xi^{-1/2}$, which is skew-symmetric and invertible due to the properties of $\Omega$ (see Eq.~\reff{eq:sympl}). Accordingly, by the spectral theorem, there exists an orthogonal matrix $R \in O\lt 2n, \R \rt$ which performs the block diagonalization
		\be
			R^\intercal K R = \quadblock{0}{\bE^{-1}}{-\bE^{-1}}{0},
		\label{eq:bldiag}
		\ee		
		where $\bE^{-1}$ is a positive-definite, diagonal $n \times n$ matrix (which, as we are going to show, coincides with the one appearing in Eq.~\reff{eq:Ut}). Its positivity is guaranteed by the fact that one can always exchange a negative diagonal entry with its opposite lying in the opposite block $-\bE^{-1}$ by exchanging the two vectors identified by the corresponding row and column via an orthogonal transformation. Being positive-definite, its inverse square root $\bE^{1/2}$ exists and we can use it to define the diagonal $2n \times 2n$ matrices
		\be
			{\bf D} = \quadblock{\bE^{1/2}}{0}{0}{\bE^{1/2} }\quad \text{ and } \quad S = \Xi^{-1/2} R {\bf D}.
		\ee
		The last one is exactly the symplectic transformation we were looking for; in fact, 
		\be
			S^\intercal \Omega S \eq    \lt {\bf D} R^\intercal \Xi^{-1/2}  \rt \, \Omega  \lt \Xi^{-1/2} R {\bf D} \rt \eq {\bf D} R^\intercal K R {\bf D}    \eq    \Omega,
		\ee
		where we have applied the definition in Eq.~\reff{eq:bldiag} and used the fact that the transposed of a symplectic matrix is still symplectic and, moreover, 
		\be
			S^\intercal \Xi S = \lt   {\bf D} R^\intercal \Xi^{-\frac{1}{2}}  \rt   \Xi   \lt    \Xi^{-\frac{1}{2}} R {\bf D}  \rt   = {\bf D} R^\intercal R {\bf D} = {\bf D}^2,
		\ee
		where we used the fact that $R$ is orthogonal, i.e., $R^\intercal = R^{-1}$. Hence, we see that the Hamiltonian $H'$ in Eq.~\reff{eq:cb2} is recast into the form
		\be
			H' = \lt\vec{\rho} \rt^\intercal {\bf D}^2 \vec{\rho} - \trace{\beta} = \lt \vec{\rho} \rt^\intercal \quadblock{\bE}{0}{0}{\bE} \vec{\rho} - \suml{q>0}{} \epsilon_q 
		\ee
		which, applying the transformation $U^\dag (\cdot) U$ to retrieve the representation in terms of particle creation and annihilation operators and denoting with $\left\{ E_q/2 \right\}_q$ the spectrum of $\bE$, yields
		\be
		\begin{split}
			H' &= \vec{\eta}^\dag \quadblock{\bE}{0}{0}{\bE}  \vec{\eta} - \suml{q>0}{} \epsilon_q = \\
			& = \suml{q>0}{} \frac{E_q}{2} \lt \eta_q^\dag \eta_q + \eta_q \eta_q^\dag  \rt -  \suml{ q > 0 }{} \epsilon_q,
		\end{split}		
		\ee
		which corresponds to Eq.~\reff{eq:freeH'} with $\mathcal{C} = \sum_{q>0} \lt E_q/2 - \epsilon_q \rt$.

		\end{subsection}

	\end{section}


\bibliography{biblio}{}

\end{document}